\input harvmac

%
%
\message{S-Tables Macro v1.0, ACS, TAMU (RANHELP@VENUS.TAMU.EDU)}
%
%
\newhelp\stablestylehelp{You must choose a style between 0 and 3.}%
\newhelp\stablelinehelp{You
should not use special hrules when stretching
a table.}%
\newhelp\stablesmultiplehelp{You have tried to place an S-Table
inside another
S-Table.  I would recommend not going on.}%
%
%
\newdimen\stablesthinline
\stablesthinline=0.4pt
\newdimen\stablesthickline
\stablesthickline=1pt
%
%
\newif\ifstablesborderthin
\stablesborderthinfalse
\newif\ifstablesinternalthin
\stablesinternalthintrue
\newif\ifstablesomit
\newif\ifstablemode
\newif\ifstablesright
\stablesrightfalse
%
%
\newdimen\stablesbaselineskip
\newdimen\stableslineskip
\newdimen\stableslineskiplimit
%
%
\newcount\stablesmode
\newcount\stableslines
\newcount\stablestemp
\stablestemp=3
\newcount\stablescount
\stablescount=0
\newcount\stableslinet
\stableslinet=0
%
%
%
\newcount\stablestyle
\stablestyle=0
%
%
\def\stablesleft{\quad\hfil}%
\def\stablesright{\hfil\quad}%
%
%
\catcode`\|=\active%
%
%
\newcount\stablestrutsize
\newbox\stablestrutbox
\setbox\stablestrutbox=\hbox{\vrule height10pt depth5pt width0pt}
\def\stablestrut{\relax\ifmmode%
                         \copy\stablestrutbox%
                       \else%
                         \unhcopy\stablestrutbox%
                       \fi}%
%
%
\newdimen\stablesborderwidth
\newdimen\stablesinternalwidth
\newdimen\stablesdummy
\newcount\stablesdummyc
\newif\ifstablesin
\stablesinfalse
%
%
\def\begintable{\stablestart%
  \stablemodetrue%
  \stablesadj%
  \halign%
  \stablesdef}%
\def\stablesadj{%
  \ifcase\stablestyle%
    \hbox to \hsize\bgroup\hss\vbox\bgroup%
  \or%
    \hbox to \hsize\bgroup\vbox\bgroup%
  \or%
    \hbox to \hsize\bgroup\hss\vbox\bgroup%
  \or%
    \hbox\bgroup\vbox\bgroup%
  \else%
    \errhelp=\stablestylehelp%
    \errmessage{Invalid style selected, using default}%
    \hbox to \hsize\bgroup\hss\vbox\bgroup%
  \fi}%
\def\stablesend{\egroup%
  \ifcase\stablestyle%
    \hss\egroup%
  \or%
    \hss\egroup%
  \or%
    \egroup%
  \or%
    \egroup%
  \else%
    \hss\egroup%
  \fi}%
\def\stablestart{%
  \ifstablesin%
    \errhelp=\stablesmultiplehelp%
    \errmessage{An S-Table cannot be placed within an S-Table!}%
  \fi
  \global\stablesintrue%
  \global\advance\stablescount by 1%
  \message{<S-Tables Generating Table \number\stablescount}%
  \begingroup%
  \stablestrutsize=\ht\stablestrutbox%
  \advance\stablestrutsize by \dp\stablestrutbox%
  \ifstablesborderthin%
    \stablesborderwidth=\stablesthinline%
  \else%
    \stablesborderwidth=\stablesthickline%
  \fi%
  \ifstablesinternalthin%
    \stablesinternalwidth=\stablesthinline%
  \else%
    \stablesinternalwidth=\stablesthickline%
  \fi%
  \tabskip=0pt%
  \stablesbaselineskip=\baselineskip%
  \stableslineskip=\lineskip%
  \stableslineskiplimit=\lineskiplimit%
  \offinterlineskip%
  \def\borderrule{\vrule width \stablesborderwidth}%
  \def\internalrule{\vrule width \stablesinternalwidth}%
  \def\thinline{\noalign{\hrule height \stablesthinline}}%
  \def\thickline{\noalign{\hrule height \stablesthickline}}%
  \def\trule{\omit\leaders\hrule height \stablesthinline\hfill}%
  \def\ttrule{\omit\leaders\hrule height \stablesthickline\hfill}%
  \def\tttrule##1{\omit\leaders\hrule height ##1\hfill}%
  \def\stablesel{&\omit\global\stablesmode=0%
    \global\advance\stableslines by 1\borderrule\hfil\cr}%
  \def\el{\stablesel&}%
  \def\elt{\stablesel\thinline&}%
  \def\eltt{\stablesel\thickline&}%
  \def\elttt##1{\stablesel\noalign{\hrule height ##1}&}%
  \def\elspec{&\omit\hfil\borderrule\cr\omit\borderrule&%
              \ifstablemode%
              \else%
                \errhelp=\stablelinehelp%
                \errmessage{Special ruling will not display properly}%
              \fi}%
  \def\stmultispan##1{\mscount=##1 \loop\ifnum\mscount>3
\stspan\repeat}%
  \def\stspan{\span\omit \advance\mscount by -1}%
  \def\multicolumn##1{\omit\multiply\stablestemp by ##1%
     \stmultispan{\stablestemp}%
     \advance\stablesmode by ##1%
     \advance\stablesmode by -1%
     \stablestemp=3}%
  \def\multirow##1{\stablesdummyc=##1\parindent=0pt\setbox0\hbox\bgroup%
    \aftergroup\emultirow\let\temp=}
  \def\emultirow{\setbox1\vbox to\stablesdummyc\stablestrutsize%
    {\hsize\wd0\vfil\box0\vfil}%
    \ht1=\ht\stablestrutbox%
    \dp1=\dp\stablestrutbox%
    \box1}%

\def\stpar##1{\vtop\bgroup\hsize ##1%
     \baselineskip=\stablesbaselineskip%
     \lineskip=\stableslineskip%

\lineskiplimit=\stableslineskiplimit\bgroup\aftergroup\estpar\let\temp=}%
  \def\estpar{\vskip 6pt\egroup}%
  \def\stparrow##1##2{\stablesdummy=##2%
     \setbox0=\vtop to ##1\stablestrutsize\bgroup%
     \hsize\stablesdummy%
     \baselineskip=\stablesbaselineskip%
     \lineskip=\stableslineskip%
     \lineskiplimit=\stableslineskiplimit%
     \bgroup\vfil\aftergroup\estparrow%
     \let\temp=}%
  \def\estparrow{\vfil\egroup%
     \ht0=\ht\stablestrutbox%
     \dp0=\dp\stablestrutbox%
     \wd0=\stablesdummy%
     \box0}%
  \def|{\global\advance\stablesmode by 1&&&}%
  \def\|{\global\advance\stablesmode by 1&\omit\vrule width 0pt%
         \hfil&&}%
  \def\vt{\global\advance\stablesmode by 1&\omit\vrule width
\stablesthinline%
          \hfil&&}%
  \def\vtt{\global\advance\stablesmode by 1&\omit\vrule width
\stablesthickline%
          \hfil&&}%
  \def\vttt##1{\global\advance\stablesmode by 1&\omit\vrule width ##1%
          \hfil&&}%
  \def\vtr{\global\advance\stablesmode by 1&\omit\hfil\vrule width%
           \stablesthinline&&}%
  \def\vttr{\global\advance\stablesmode by 1&\omit\hfil\vrule width%
            \stablesthickline&&}%
  \def\vtttr##1{\global\advance\stablesmode by 1&\omit\hfil\vrule
width ##1&&}%
  \stableslines=0%
  \stablesomitfalse}
\def\stablesdef{\bgroup\stablestrut\borderrule##\tabskip=0pt plus 1fil%
  &\stablesleft##\stablesright%
  &##\ifstablesright\hfill\fi\internalrule\ifstablesright\else\hfill\fi%
  \tabskip 0pt&&##\hfil\tabskip=0pt plus 1fil%
  &\stablesleft##\stablesright%
  &##\ifstablesright\hfill\fi\internalrule\ifstablesright\else\hfill\fi%
  \tabskip=0pt\cr%
  \ifstablesborderthin%
    \thinline%
  \else%
    \thickline%
  \fi&%
}%
\def\endtable{\advance\stableslines by 1\advance\stablesmode by 1%
   \message{- Rows: \number\stableslines, Columns:
\number\stablesmode>}%
   \stablesel%
   \ifstablesborderthin%
     \thinline%
   \else%
     \thickline%
   \fi%
   \egroup\stablesend%
\endgroup%
\global\stablesinfalse}
%

\baselineskip=.55truecm
\Title{\vbox{\hbox{HUTP--96/A054}\hbox{IASSNS-HEP-96/119}\hbox{hep-th/9701165}}}{\vbox{\centerline{On Four-Dimensional Compactifications}
\vskip .1in
\centerline{of   F-Theory}}}
\vskip .05in
\centerline{\sl Michael Bershadsky$^{\natural}$,
Andrei  Johansen$^{\natural}$,
Tony Pantev$^{\forall}$ and
Vladimir Sadov$^{\sharp}$ }
\vskip .2in
\centerline{\it $^{\natural}$ Lyman Laboratory of Physics, Harvard
University}
\centerline{\it Cambridge, MA 02138, USA}
\vskip .2in
\centerline{\it $^{\forall}$ Department of Mathematics, Massachusetts
Institute of Technology}
\centerline{\it Cambridge, MA 02138, USA}
\vskip .2in
\centerline{\it $^{\sharp}$ Institute for
Advanced Study}
\centerline{\it Princeton, NJ 08840, USA}

\vskip .4in
\centerline{ABSTRACT}
\vskip .2in
\noindent
Branches of moduli space of F-theory in four dimensions are investigated.
The transition between two  branches is described as a
3-brane--instanton transition on a 7-brane. A dual heterotic picture of
the transition
is presented and the F-theory --- heterotic theory map is given. The
F-theory
data --- complex structure of the Calabi-Yau fourfold and the
instanton bundle
on the 7-brane is mapped to the heterotic bundle on the elliptic
Calabi-Yau
threefold $CY_3$. The full moduli space  has a web structure
which is also
found in the moduli space of semi-stable bundles on $CY_3$.  Matter
content of
the four-dimensional theory is discussed in both F-theory and
heterotic
theory descriptions.

\vskip .2in

\Date{}

\vfill\eject


\def\P{{\bf P}}
\def\F{{\bf F}}
\def\V{{\cal V}}
\def\S{{\Sigma}}
\def\Pm1{{\bf P}^1}
\def\Pt1{${\bf P}^1$}
\def\unc{\underline{c}}
\def\W{{\cal W}}

\newsec{Introduction}

Great progress has been made recently in our  understanding of
six-dimensional
compactifications of F-theory on elliptic Calabi-Yau threefolds
\ref\MV{D. Morrison and C. Vafa, {\it Compactifications of F-theory on
Calabi-Yau
threefolds-
I, II}, Nucl. Phys. {\bf B 473} (1996) 74; {\it ibid}  {\bf B 476} (1996)
437.}.
The structure of the
singular locus
of elliptic fibration encodes the information about both the
enhanced gauge
symmetries and the matter contents of F-theory compactification
\MV \ref\kucha{M. Bershadsky, K. Intriligator,
S. Kachru, D. Morrison, V. Sadov, and C. Vafa, {\it Geometric
Singularities and Enhanced Gauge Symmetries},
Nucl. Phys. {\bf B 481} (1996) 215.}.
F-theory also provides us with a powerful
tool in studying the nonperturbative aspects of heterotic string
compactifications, in the case when the heterotic dual exists.

In this paper we focus mainly on four-dimensional compactifications of
F-theory. Four-dimensional compactifications appear to be very different
 from  six-dimensional ones. First of all, these compactifications
generically have
a 3-brane anomaly \ref\wit {C. Vafa and E. Witten,
Nucl. Phys. {\bf B447} (1995) 261.}\ref\sis{K. Becker and M.
Becker, Nucl.
Phys. {\bf B477} (1996) 155.}\ref\wvs{S. Sethi, C. Vafa and
E. Witten, Nucl. Phys. {\bf B480} (1996) 213.}.  The
RR 4-form has an uncompensated 3-brane charge
\eqn\an{\alpha={1\over 24}{\chi (CY_4)}, }
where $\chi$ is the
Euler number of Calabi-Yau fourfold.  In order to cancel this
anomaly  one
can  insert an appropriate number of 3-branes
\foot{In this paper we will not discuss effects of possible
nontrivial discrete background 3-brane fluxes
\ref\fluxes{E. Witten, {\it On flux quantization in M-theory and the
effective action}, hep-th/9609122.}.}.
When F-theory has a heterotic dual these 3-branes should correspond
to the
heterotic 5-branes \ref\witun{E. Witten, unpublished.}\ref\bds{
T. Banks, M. Douglas and N. Seiberg, Phys. Lett.
{\bf B387} (1996) 278.}. Therefore, one novel feature is
that we have to learn how to
deal with 3-(5-) branes.

Another novelty is that
on the compact part of the world-volume of the 7-brane, one can turn on
 the gauge field background \ref\wdyn{E. Witten, Nucl. Phys.
{\bf B460} (1996) 335.}  with a nonzero instanton number. Only
when the background is trivial the
four-dimensional gauge group is  the one prescribed by the
singularities of the elliptic fibration. Any nontrivial background
breaks the
gauge group to a smaller one.  Also, in the presence of nontrivial
background, the anomaly
counting \an\ changes: $\alpha=\chi/24 - k$, where $k$
is the total number of instantons inside the 7-branes.

The properties of the four-dimensional N=1 supersymmetric field
theories are
determined by the configuration of 7-branes and 3-branes that intersect
over a common flat $R^{3,1}$.  The gauge groups come from both
7-branes and 3-branes.
Many ways to distribute the anomaly $\alpha$ between
3-branes and instantons give rise to many
branches of N=1 four-dimensional theory. These branches and
transitions between them have a nice
interpretation in terms of D-brane physics.
For example, a single 3-brane
produces a $U(1)$ factor in the full gauge group of the four-dimensional
theory. It also contributes by $1$ to cancellation of $\alpha$.
The position of the 3-branes inside the base of the elliptic
fibration parameterizes the moduli space of the $U(1)$ theory.
In particular it  determines \bds \ref\soloma{A. Sen, {\it BPS
States on a Three Brane Probe}, hep-th/9608005.} the masses of the chiral
superfields
coming from strings connecting the 3-brane with  7-branes.
When the 3-brane approaches a 7-brane, some of these
fields become massless. At this very moment a transition to the
Higgs branch
becomes possible {\it if} some conditions are
satisfied.
 On the Higgs branch  a 3-brane ``dissolves'' into a finite size
instanton of the nonabelian gauge group \ref\ww{E. Witten,
{\it Small Instantons in String Theory},
Nucl. Phys.
{\bf B460} (1996) 541.}\ref\doug{M. Douglas, {\it Gauge Fields and
D-branes}, hep-th/9604198.} and the 3-brane $U(1)$ gauge group  
disappears.
If the appropriate conditions are not satisfied, a superpotential gets
generated that prevents a theory from developing the Higgs branch.

The pure Higgs branch corresponds to the situation when all 3-branes are
replaced by the instantons and the anomaly is cancelled by the
nonabelian gauge
field.  The mixed branches are the ones with both 3-branes and
instantons.

If an F-theory compactification has a heterotic dual, this variety
of branches
finds its counterpart in the variety of branches of the moduli
space of bundles
on Calabi-Yau threefolds. In this paper we mainly consider the $SU(n)$
vector bundles. Very much unlike the situation with  bundles on
$K3$, the
moduli space ${\cal M}_{CY_3}(n, c_2, c_3)$ of bundles with the
fixed rank and
Chern classes can have many irreducible components with different
dimensions.
To describe bundles on {\it elliptic} Calabi-Yau's we will use a very
useful tool --- the theory of spectral covers.
This mathematical construction is well-known in the context of Hitchin
systems
\ref\spectralth{See, for example, R. Donagi, {\it Spectral covers}, in
Current topics in complex algebraic geometry (Berkeley, CA,
1992/93), 65-86, MSRI Publ. 28 and R. Donagi and E. Markman
{\it Spectral covers, algebraically completely integrable,
Hamiltonian systems, and moduli of bundles} in Integrable systems and
quantum groups (Montecatini Terme,
1993), 1-119, LNM 1620, and references therein.}.
Its applications to the heterotic
string compactifications are developed in ref.
\ref\WM{R.~Friedman, J.~Morgan and E.~Witten,
{\it Vector Bundles and F Theory}, hep-th/9701162.}.
 We will use a slightly different formulation of
this approach
which is suitable to deal with different components of the moduli space.

Some aspects of Calabi-Yau fourfold compactifications of F-theory
have been considered recently
in refs.~\ref\brunner{I. Brunner and R. Schimmrigk, {\it F-Theory on
Calabi-Yau Fourfolds},
Phys. Lett. {\bf B387} (1996) 750.}\ref\bianchi{M. Bianchi, S.  
Ferrara, G.
Pradisi,
A. Sagnotti and Ya. S. Stanev,
{\it Twelve-Dimensional Aspects of Four-Dimensional N=1
Type I Vacua}, Phys. Lett. {\bf B387} (1996) 64.}\ref\kachru{S.  
Kachru and
E. Silverstein, {\it Singularities, Gauge Dynamics, and
Nonperturbative Superpotentials in
String Theory}, hep-th/9608194.}\ref\mayr{P. Mayr, {\it Mirror Symmetry,
N=1 Superpotentials and Tensionless Strings on Calabi-Yau
Fourfolds}, hep-th/9610162.}\ref\bru{ I. Brunner,  M. Lynker,  R.  
Schimmrigk,
{\it Unification of M- and F- Theory Calabi-Yau Fourfold Vacua},
hep-th/9610195.}\ref\katzvafa{S. Katz and C. Vafa,
{\it Geometric Engineering of N=1 Quantum Field Theories},
hep-th/9611090.}\ref\bjsv{M. Bershadsky, A. Johansen,
T. Pantev, V. Sadov and C.  Vafa, {\it F-theory, Geometric  
Engineering and
N=1 Dualities}, hep-th/9612052.}\ref\zw{C. Vafa and B. Zwiebach, {\it N=1
Dualities of SO and USp Gauge Theories
and T-Duality of String Theory}, hep-th/9701015.}\ref\klemm{A.
Klemm, B. Lian, S.-S. Roan and S.-T. Yau,
{\it Calabi-Yau fourfolds for M- and F-theory compactifications},
hep-th/9701023.}.

\bigskip

The plan of this paper is as follows.
In section 2 we review how one can describe a six-dimensional
compactification using the adiabatic arguments of
\ref\vw{C. Vafa and E. Witten,
{\it Dual String Pairs With N=1 And N=2 Supersymmetry In Four  
Dimensions},
hep-th/9507050.}.  One can  fiber eight-dimensional theory data (it
could be either F-theory or its heterotic dual) over a ${\P}^1$.
F-theory is defined on elliptic Calabi-Yau threefold which is also a
K3 fibration over \Pt1.  The base of the elliptic fibration is a
rational ruled surface ${\bf F}_n$.
The heterotic dual is characterized by the distribution of instanton
numbers (12+n, 12-n) between two $E_8$'s.
Using the adiabatic arguments we give a nice geometric description
of vector bundles on elliptic K3.
This description is nothing else but the spectral cover
theory \WM\ for $K3$. It allows us to reformulate various
statements about F-theory
-- heterotic duality in a way preparing the reader for the more
complex story awaiting him or her in four dimensions.

In section 3  we push the adiabatic argument further, down to four
dimensions.
In doing this we compactify heterotic string on the elliptic Calabi-Yau
threefold.  The F-theory dual is defined on Calabi-Yau fourfold which
at the same time is a K3 fibration. The base of this elliptic CY
fourfold is  a $\P^1$ bundle over $\F_n$ which we call a
generalized Hirzebruch variety
$\F_{nmk}$, where the indices
$m,~k$ indicate how the sphere $\P^1$ is fibered over $\F_n$. We
discuss the
spectral theory of vector bundles on elliptic Calabi-Yau threefolds
and use it to
describe various branches of their moduli spaces.

In section 4 we discuss various branches of the moduli spaces of the
F-theory compactifications and  relations between them.
The full moduli space is a huge web  which
includes the
moduli of the Calabi-Yau fourfold, gauge fields inside 7-branes and
positions of 3-branes.
All these moduli
spaces are interrelated and the transition
points correspond to the singularities in the moduli space.  In the same
context we discuss the relation between the 3-branes and the
heterotic 5-branes. We present a map
identifying the moduli spaces of F-theory and heterotic compactification.

Section 5 deals with the moduli appearing in the transition from
one branch to
another. We compute the number of such moduli and explain their
meaning both in
F-theory and in heterotic string theory. We also address the
general question
 about the matter content of F-theory.

In the Appendix we present various mathematical statements
used in this
paper. Detailed proofs of some of the theorems will appear
elsewhere.

\newsec{Review of F-theory -- heterotic duality in six dimensions}

\subsec{Heterotic string}

Here we briefly review F-theory -- heterotic string duality in 6
dimensions but from a slightly different angle. This point of view allows
us to generalize various six-dimensional results to four dimensions.

We first consider heterotic theory compactified on two-dimensional torus
$T^{2}$
\ref\nar{K. Narain, Phys. Lett. {\bf B169} (1986) 41;
K. Narain, H. Sarmadi and E. Witten, Nucl. Phys. {\bf B 279} (1991)  
526.}.
The heterotic string theory in 8 dimensions is uniquely defined
by
specifying a complex and K\"{a}hler structure on $T^{2}$ and a
holomorphic
$E_8 \times E_8$ bundle on $T^{2}$.
The moduli space of $E_8 \times E_8$ bundles on $T^2$ is the same as the
moduli
of representations of $\pi_{1}(T^2)$ into $E_8 \times E_8$. The later is
easily identified with $({\rm Hom}(\pi_{1}(T^2),H))^{W}$ where $H \subset
E_8 \times E_8$ is a Cartan torus and $W$ is the Weyl group.
Alternatively
this moduli space can be written as
\eqn\modd{({\bf C}\otimes \Gamma_{E_{8}\times
E_{8}})^{W}/H_{1}(T^{2},{\bf Z})}
where $\Gamma_{E_{8}\times E_{8}}$ is the
co-root lattice of $H$ and we have realized $T^{2}$ as the quotient
${\bf C}/H_{1}(T^{2},{\bf Z})$.

Let us vary the eight-dimensional heterotic data over an additional  
$\P^{1}$
so that the family of complex tori  fits into an elliptic $K3$.
In order to formulate a heterotic string theory on this $K3$, we  
will also
require that the $E_{8}\times E_{8}$ bundles on the fibers fit into a
global
holomorphic bundle, which we denote by $\V=V_1 \times V_2$.

It would be desirable to have a description of $\V$ in terms of
information
concentrated along the fibers and the base of the $K3$ surface. By
restricting $\V$ on the fibers and on the zero section of $K3 \rightarrow
\P^{1}$ we obtain a family of flat bundles on the fibers and a flat
bundle
on the base\foot{By a flat bundle on a Riemann surface we mean a
principal
bundle admitting a holomorphic flat connection. In particular, the
bundles
$\V \mid _{T^{2}}$ can be flat without the restriction of the instanton
connection
on $\V$ being flat.}. Naively one would expect that this collection of
data
suffices to reconstruct $\V$. However,
the information captured by these
restrictions is incomplete and does not reflect the monodromy of the
connection matrices on $\V \mid_{T^{2}}$ when we go around some special
points on
the base.

The precise relation between the  family $\{\V \mid_{T^2}\}$ of  
flat bundles
and $\V$  can be
made explicit. To simplify the exposition we will discuss only the
$SU(r)$ and $SO(n)$ vector bundles.  We start with the
$SU(r)$ bundles.
The moduli space of $SU(r)$  flat bundles over the torus $T^2$
is a complex projective space  $\P^{r-1}$.
One can think about a point in $\P^{r-1}$ as a set of $r$ points
$(x_1,\ldots ,x_r) $ in
the dual torus $\check{T}^2$,
subject to a constraint $\sum x_i=0$. The set $(x_1,\ldots ,x_r)$
  parameterizes the $SU(r)$
Wilson line around $T^2$.

When we have a family $\{\V \mid_{T^2}\}$ of flat
 $SU(r)$
bundles parameterized by the projective line $\P^1$, the above  
construction
produces a {\it spectral curve} $\Sigma \subset {\check{K3}}$.   
{\it The dual}
$K3$ denoted by ${\check{K3}}$ is the elliptic fibration over \Pt1\  
obtained
from the original elliptic $K3$ by replacing each fiber $T^2\rightarrow
\check{T}^2$.
The spectral curve $\S$ projects onto \Pt1\ so that the preimage of
a point $p\in \P^1$ is the set $(x_1,\ldots ,x_r)$ corresponding to
the restriction $\V\mid_{T^2}$ of $\V$ to the fiber $T^2$ over $p$.
In general the spectral curve consists of several irreducible components
$\Sigma_i$ with multiplicities $r_i$. Multiplicity $r_i>1$ means  
that for any
$p\in \P^1$, the line bundle $x_i\in \check{T^2}$ can be found  
$r_i$ times
in the decomposition of $\V\mid_{T^2}$.
Each curve   $\Sigma_i$ may
cover the base several times $n_i$.
The class of the spectral curve $\Sigma$ is
given by
\eqn\speccur{\S=r S+k F~,}
where $S$ is the zero section and $F$ is the fiber
 and $r=\sum r_i n_i$. The coefficient $k$ should be
identified
with $c_2(V)$.

The concept  of  spectral curve is very useful
when we compare the heterotic compactification on $K3$ with the dual
F-theory
compactification on a Calabi-Yau threefold $CY_3$. What the
adiabatic argument  \vw\ essentially tells us is that
the spectral curve $\S$ (together with $K3$) {\it determines the complex
structure of the F-theory  Calabi-Yau
threefold}.  To be precise, the complex moduli space of elliptic
Calabi-Yau threefolds ${\cal M}^{(3)}$ can be represented as
a bundle
$({\cal M}^{(3)} \rightarrow {\cal M}_{K3})$, where the fiber
${\cal M}_{\S}$ is the moduli space of the spectral surface $\S$ and
${\cal M}_{K3}$ is the moduli space of complex structures of $K3$. The
number
of complex deformations of the singular locus of Calabi-Yau threefold
coincides with the number of complex deformations of the spectral
curve, which
is
equal to its arithmetic genus (see Appendix)
\eqn\gen{g(\S)={1 \over 2} \S^2+1=r k- r^2+1~.}
In the generic situation the spectral curve $\S$ consists of two
curves with
multiplicities one, each corresponding
to the bundles $V_{1,2}$.

The information encoded in the spectral curve $\Sigma(\V)$ is  not
sufficient to reconstruct the bundle $\V$ on $K3$. The moduli space  
of vector
bundles on K3 is
a hyperk\"ahler variety and  ${\cal M}_\S$ is not hyperk\"ahler  
(this variety
is a projective space). Also the complex dimension of the moduli space of
vector bundles is twice the dimension of ${\cal M}_\S$. In fact, we have
already encountered this situation in \ref\topd{M.~Bershadsky,  
V.~Sadov and
C.~Vafa,
 {\it D-branes and Topological Field Theories}, Nucl.~Phys.~{\bf  
B463} 420.}
 in discussing the supersymmetric cycles in $K3$; so here we may
simply borrow the result.
A reader will find a more mathematically rigorous approach in
the Appendix.
It turns out that to recover the full moduli space
the spectral curve $\Sigma$ should
be equipped with a
line bundle $L$ of degree ${\rm deg} L= - (r+g-1)$.
The pushforward of this line bundle
on the base yields a vector bundle of rank $r$ which coincides with the
restriction of $\V_S$ to the zero section.

Proposition 1 stated in the Appendix reads that a pair $(\Sigma ,L)$
uniquely determines the vector bundle $V$ with the trivial  first  
Chern class
and the second Chern class equal to $c_2(V)=k$.
It is quite remarkable that the genus of the spectral curve is equal
to half of
the dimension of the
moduli space of the vector bundle.

The number of matter multiplets can also be computed in terms of spectral
curves.  Suppose that
the
bundle $V$ is a $G$-bundle where $G$ is the broken subgroup
of $E_8$.
Let $H$ be the unbroken subgroup so that
$H \times G \subset E_8$. Let $S_i$ be the representations of $G$,
entering
into the decomposition ${\bf 248}=\bigoplus (S_i \otimes R_i )$, where
$R_i$ are the representations of the unbroken group $H$.
With each representation $S_i(V)$ one can associate a spectral curve
$\Sigma_i$.
It follows from the index theorem \WM\ that the number of matter
multiplets  in
representation $R_i$ of $H$
equals
\eqn\numm{N(R_i )=\S_i \cdot S ~.}
It is also quite interesting to consider the case of $SO(n)$
bundles\foot{These $SO(n)$ bundles are associated with $Spin(n)$ bundles,
imbedded into $E_8$.}. One can describe
$SO(n)$ flat bundles on a torus in terms of $n$ points, invariant under
the ${\bf Z}_2$ involution.
The involution flips the sign of the flat coordinate along the torus
(${\bf Z}_2:z \rightarrow -z$). This description gives rise to a
${\bf Z}_2$ invariant spectral curve
$\S$ which covers the base $n$ times.
${\bf Z}_2$ involution permutes the sheets when $n$ is even. In the case
when $n$ is odd the spectral curve is reducible $\S= S +
\S_{n-1}$ with the zero section $S$ being fixed by the involution.
The spectral curve $\S$ should
be equipped
with {\it anti-invariant} line bundle.
The class of the spectral curve is equal to
\eqn\scso{\S=n S+ 2c_2(V) F.}
Let us count the number of relevant deformations of the spectral curve
$\S$.
Let $H^{0}(N_{\S})$ be the space of global sections of the normal bundle
to
$\S$. Since the involution preserves $\S$ this space can be  
decomposed into
the sum $H^{0}(N_{\S}) = H_{+}^{0}(N_{\S})\oplus  
H_{-}^{0}(N_{\S})$, where
$H_{\pm}^{0}(N_{\S})$ are the invariant (anti-invariant subspaces). The
deformations we are interested in are the ones preserving the action of
the involution and thus their number is equal to $H_{+}^{0}(N_{\S})$ for
an even $n$ and to $H_{+}^{0}(N_{\S_{n-1}}(S))$ for an odd $n$.
Alternatively
in terms of the line bundle ${\cal O}(n S+ 2c_2(V) F)$ with its natural
${\bf Z}_{2}$ action, the number of relevant deformations of $\S$  
is given
by $\dim H^{0}_{-}(K3, {\cal O}(n S+ 2c_2(V) F))$ (regardless of the
parity
of $n$).
If we consider the natural
projection\foot{The quotient of $K3$ by ${\bf Z}_{2}$ is a Hirzebruch
surface ${\bf F}_{n}$. The zero section $S$ of $K3$ is mapped to the
infinity section $\underline{S}$ of ${\bf F}_{n}$. Computing the
self-intersections we get $-2 = S^{2} = (1/2)\underline{S}^{2} = -n/2$.
Therefore $n = 4$.} $K3 \rightarrow \F_4 = K3/{\bf Z}_{2}$, then
$n S + 2c_{2}(V) F$
is the preimage of the ${\bf Q}$-class
$\sigma = (n/2)\underline{S} + 2c_{2}(V)\underline{F}$, where
$\underline{S}$
and $\underline{F}$ are the infinity section and the fiber of ${\bf  
F}_{4}$,
respectively.
Now the dimension of $H^{0}_{-}(K3, {\cal O}(n S+ 2c_2(V)
F))$
can be easily computed from the Lefschetz fixed point formula and
is equal to
\eqn\defso{{\rm dim} H^{0}_{-}(K3, {\cal O}(n S+ 2c_2(V) F)) =
{1 \over 2}(\sigma^2 -c_1 \sigma)=c_2
(V) (n-2)-n(n-1)/2 ~.}
Again, it is quite remarkable that ${\rm dim} H^{0}_{-}(K3, {\cal
O}(n S+ 2c_2(V) F)) $ coincides
with half of the dimension of the $SO(n)$ instanton moduli space. The
dimension of the moduli space of anti-invariant line bundles also
coincides with \defso, which is just a consequence of the fact that the
full moduli space is hyperk\"ahler.

It is clear from the construction that the spectral curve $\Sigma$
(collection of curves $\{ \S_i \}$ with multiplicities $r_i$) encodes the
information about the Wilson lines along the fiber.
Consider the case when several components of the spectral curve $\Sigma$,
say $\S'$ and $\S''$ degenerate to the one component with nontrivial
multiplicity (for  example $ \S' + \S'' \rightarrow 2 \S$).
As a result of this degeneration two line bundles $L'$ and $L''$ are
combined into a rank $2$ bundle.
In general, the degeneration of several  components of
the spectral curve $\S_i$ to a multiple one of the form $n \S$ yields
a more complicated object, e.g. a rank $n$
vector bundle on $\S$.
In all these cases various
Wilson lines get aligned and one should expect gauge
symmetry enhancement.
In fact, this mechanism looks very similar to the one responsible for the
appearance of enhanced gauge symmetry when several parallel D-branes come
together \wdyn.
Comparing with F-theory side we can identify each
degeneration of the
spectral curve as some degeneration of the discriminant locus when
several D-branes come together.
We will discuss this phenomenon in detail in the
context of four-dimensional F-theory compactifications.

\subsec{F-theory}

F-theory in 8 dimensions is defined on an elliptic $K3$.
The moduli space of
elliptic $K3$ surfaces is known to be the quotient of the symmetric
space $SO(2,18)/SO(2)\times SO(18)$ by a discrete group
(see  \ref\wvar{E. Witten, Nucl. Phys. {\bf B443} (1995)
85.}\ref\aspin{P.
Aspinwall and D. Morrison, Phys. Lett. { \bf B355} (1995)
141.}  and references therein). It is a little bit
better to think of this space as the bounded symmetric domain
\eqn\tei{
{\cal D} = \{ w \in
\P((\Gamma_{E_{8}\times E_{8}}
 \oplus \sigma \oplus \sigma) \otimes {\bf C}) \mid
\langle w, w \rangle = 0, \langle w , \bar{w} \rangle > 0 \}.
}
The universal cover
$(\Gamma_{E_{8}\times E_{8}}\otimes {\bf C})\times SO(2,2)/(SO(2)\times
SO(2))$ of the moduli
space of the heterotic string in 8 dimensions can be identified with
${\cal D}$.
Therefore every variation of eight-dimensional heterotic data will  
produce
a family of elliptic $K3$ surfaces. If the variation leads to a
six-dimensional
heterotic theory then the corresponding family of
elliptic $K3$'s should fit into a
Calabi-Yau threefold $CY_3$.
Compactifying F-theory on this Calabi-Yau
threefold, one gets F-theory in 6 dimensions, which is dual to
the heterotic theory on $K3$.
The
complex structure of Calabi-Yau threefold is determined by the
family of the heterotic data on the fiber $T^2$ which varies over \Pt1.
Therefore only a part
of the full information about
the vector bundle $\V$ on $K3$ encodes the threefold $CY_3$.
This partial information is a family of Wilson lines $\{\V\mid
_{T^2}\}$ for every
fiber or equivalently the spectral curve. To sum up, the elliptic
fibration on
the F-theory side is determined completely by  the spectral curve
$\S$ together with heterotic $K3$.

Now let us discuss the r\^ole of the line  bundle $L$ on $\S$, which is
necessary to reconstruct $\V$.
The parameters in the polynomials in Weierstrass form
correspond to complex scalar fields of the
resulting $N=1$ six-dimensional theory, each making half of
hypermultiplet \MV. The other half of the
hypermultiplets seem to be  missing.
To recover the missing complex
parameters, one
has to take into account that each 7-brane is equipped with $U(1)$ gauge
field.
Upon compactification down to six dimensions,
exactly two components of the vector field
become scalars and can be rearranged into one complex scalar field
completing the hypermultiplet.
As we mentioned before, the spectral
curve $\S$
encodes the discriminant locus (locations of 7-branes).
The  bundle $L$ on the spectral curve $\S$ completes the data so that the
full moduli space is hyperk\"ahler. Therefore it is clear that
the line bundle $L$ encodes information about the gauge bundle
inside the 7-brane.
To be more precise, the moduli space of the gauge
fields
inside 7-brane coincides with the moduli space
of bundles on the spectral curve $\S$.

Consider the simple example of
heterotic theory with $\V=SU(n')\times SU(n'') \subset E_8 \times E_8$
bundle. Suppose that the  unbroken gauge group is $G' \times G''$.
The spectral curve $\S$ consists of at least two curves $\S'$,
$\S''$, each covering the base $n'$ and $n''$ times.
In the F-theory the discriminant locus in general consists of three
7-branes:
$D'$ and $D''$ with $G'$ and $G''$ singularities and $D_0$ with
generic $I_1$
singularity.
It is easy to check that the number of deformations preserving
 $G' \times G''$ locus matches exactly the number of  deformations
of the
curves $\S'$ and $\S''$ inside $K3$.  We will discuss similar counting in
full generality in the case of four-dimensional compactifications. The
$U(1)$ gauge bundle inside $D_0$  is determined by the line bundles on
$\S'$ and $\S''$.

\newsec{Examples of F-theory -- heterotic duality in four dimensions}

\subsec{Vector bundles and heterotic compactification}

To describe the four-dimensional compactifications we fiber
eight-dimensional data over a two-dimensional complex base  
$B_H=\F_n$.  Suppose
that eight-dimensional
data fits into an elliptic Calabi-Yau threefold $CY_3$ and a   
vector bundle
$\V=V_1\oplus V_2$.

The description of  vector bundles on the elliptic Calabi-Yau  
threefolds  is
more involved than the analogous construction for $K3$
(one can find a mathematical
discussion in the Appendix).
The discrete invariants of a vector bundle $V$ with the trivial  
first Chern
class are the rank $r$ of the bundle,
$c_3(V)$ and the components of $c_{2}(V)$.
For the threefolds that we consider, this produces five integer  
parameters.
The vector bundle is determined by the spectral surface $\S$
in the dual Calabi-Yau $\check{CY}_3$
and  a line bundle ${\cal L}$ on a smooth model
$Y$ of the fibered product
$\S \times_{B_H} CY_3$.
More precisely
the bundle $V$ is the pushforward of
$\cal L$ from Y to  $CY_3$. For every point $p \in CY_3$
the fiber of $V$ is given by $x_1 +x_2+ \ldots + x_r$, where $x_i$ are
the fibers of
$\cal L$ over the points $p_i \in \S$, such that all $p_i$ and $p$  
project on
the same point on the base $B_H$.
Let us denote by $L$ the restriction of the
bundle ${\cal L}$ to the spectral surface $\S$.
In the case of elliptic K3 the bundle ${\cal L}$
can be uniquely recovered from $L$ by pulling back and twisting  
with a certain
fixed line bundle (see Appendix).
In contrast, in the case of Calabi-Yau threefold
there is no unique way to reconstruct ${\cal L}$ from the pullback of
$L$, because one can also twist by the multiples of the exceptional  
divisor
\foot{We do not see this ambiguity in the K3 case because the fibered
product $\S \times_{B_H} K3$ is already smooth.}. This twisting
governs the third Chern class of the bundle $V$ and has no effect on
$c_1(V)$ and $c_2(V)$.
 It follows from the
construction  (see Appendix) that all deformations of the bundle
${\cal L}$ come from the deformation of the line bundle
$L \rightarrow \S$. Therefore, to simplify the discussion
 we can pretend that the vector bundle is determined by the line  
bundle  $L$,
keeping in mind that this is literally  true only if there is a relation
between $c_3(V)$  and $c_2 (V)$.

The homological class of the spectral surface is determined by the  
rank of the
bundle and
its second Chern class
\eqn\clsche{\S=rS+ c_2(V)_{AS} A +c_2(V)_{BS} B~, }
where $r$ is the rank of the bundle and  $c_2(V)_{AS, BS}$
are the coefficients in the decomposition of $c_2$ with respect to
a basis $AB$, $AS$ and $BS$. The cycles $A,B$ and $S$ make the basis in
$H_2(CY_3)$ (see Appendix). The simplest way to derive \clsche\
(in the case $n=0$) is to restrict
the bundle to the cycles $A$ and $B$. These cycles can be  
represented by $K3$
and therefore we can use the relation \scso. In general, for  $n  
\neq 0$ the
cycle $B+{n \over 2} A$ can be represented by $K3$ and the  
coefficients in
\clsche\ are determined by restricting
the
bundle $V$ to the cycles $A$ and $B+{n \over 2} A$.

The spectral surface $\S$ encodes {\it the complex structure} of  
the Calabi-Yau
fourfold $CY_4$ in F-theory.  Let us denote by ${\cal M}_{\S}$ the  
moduli space
of complex deformation of the
spectral surface.  Then the moduli space ${\cal M}^{(4)}$
 of complex
structures of
the Calabi-Yau fourfold
can be described as
the bundle $({\cal M}^{(4)} \rightarrow {\cal M}^{(3)})$ with  a fiber
${\cal
M}_{\S}$, where
${\cal M}^{(3)}$ is the moduli space of complex structures of Calabi-Yau
threefold $CY_3$ of the heterotic compactification.

When $V$ is an $SU(r)$ vector bundle, the number of complex deformations
of the spectral surface follows from the Riemann-Roch theorem and
is equal to
\eqn\class{{\rm dim}\,H^{(2,0)}(\S)={1 \over 12 }
(2\S^3+\S \cdot c_2(T_{CY}))-1 ~.}
In the next section we show that ${\rm dim}H^{(2,0)}(\S)$ matches exactly
the number of complex deformations preserving the gauge symmetry
enhancement locus on the F-theory side. Expression \class\ computes the
number of deformations under the assumption that the
spectral surface $\S$ is generic and irreducible.
As we will see in the next section when the spectral surface degenerates
and becomes
reducible one has to compute the number of deformations {\it preserving}
the number of irreducible components.

Let us briefly describe what is going on in the case of $SO(n)$ bundles.
As in six dimensions the spectral surface should be invariant under
${\bf Z}_2$ involution. The spectral surface should also be equipped with
{\it anti-invariant} line bundle. It turns out that not every ${\bf Z}_2$
invariant spectral surface allows an {\it anti-invariant} line bundle.
To count the number of parameters of spectral surfaces admitting an
anti-invariant line bundle one needs to solve an explicit  
Noether-Lefschetz
problem.
Therefore this situation differs from the six-dimensional one.
We will discuss the details of $SO(n)$ computations in
\ref\fut{M. Bershadsky, A. Johansen, T. Pantev and V.
Sadov, work in progress.}.

The bundle $L$ on $\S$ encodes the information about {\it the gauge  
fields}
inside 7-branes.
It is convenient to break the description into different
sectors according to the different geometric behavior of the components
of the spectral surface.
For simplicity we will discuss only the two limit
cases.
First, suppose that the surface $\S'$ is a component of the  
spectral surface
$\S$ with multiplicity one.
Then the line bundle $L' =L_{\mid \S'}$ has no deformations because
$h^1(\S')=0$.
This corresponds to the fact that
the $U(1)$ gauge field inside
the 7-brane has no  extra moduli.
The situation is different when the multiplicity of $\S'$ is
$n>1$.
In this case the non-reduced surface $n\S'$ is equipped with
a sheaf $L'$ that has numerical rank one (as measured by the Hilbert
polynomial).
The space of such sheaves has several connected components
labeled by the collections of ranks and degrees of the restrictions  
of $L'$
on the infinitesimal neighborhoods of $\S' \subset \check{CY}_{3}$  
of orders
$0, 1, 2,
\ldots, n-1$.
For example we will have a component parameterizing rank $n$
vector bundles on the reduced surface $\S'$ and a component  
parameterizing all
line bundles on the full non-reduced surface $n\S'$.
In the case of a
6-dimensional compactification, when
$n\S'$ is  multiple curve sitting on a (not necessarily  
elliptically fibered)
$K3$ surface a detailed description of these components in their  
structure
can be found in \ref\del{Ron Donagi, Lawrence Ein, Robert Lazarsfeld,
{\it A non-linear deformation of the Hitchin dynamical system},
alg-geom/9504017.}.
 It is very interesting to find the branches of the
$F$-theory compactifications corresponding to these components.
The analysis
of the geometry of these branches is rather involved and  will be a  
subject
to a future investigation \fut .
For now we will examine the special case when
the sheaf $L'$ is determined by a rank $n$ vector
bundle $M$ on $\S'$, which is characterized by $c_2(M)$.
To explain how this vector bundle appears let us consider a concrete
example.
Let $V=\pi^*M$ be a vector bundle on the Calabi-Yau threefold  
$CY_3$ which is
 a pullback of the bundle $M$ on the base $B_H$;
 we denote by $\pi$ the projection $\pi: CY_3 \rightarrow B_H$.
Restricted to any fiber,
$\pi^* M\mid_{T^2}$ is a trivial rank $n$ bundle.
The
corresponding spectral surface is the zero section $S$ taken with
multiplicity $n$: $\Sigma(\pi^* M)=nS$.
The spectral bundle on $S$ is $M$ itself, considered as
a bundle on the zero section.

To appreciate the r\^ole of  bundle
$M$ on the spectral surface with multiplicity,
let us consider another example. Choose a  special vector bundle
$V=E\oplus\cdots\oplus E=E \otimes I_n$
where $E$ is any irreducible bundle and $I_n$ is a trivial vector bundle
of rank $n$.
The spectral surface of $V$ is the same as the spectral surface of $E$
 taken with multiplicity $n$.
The bundle $V$ has a large group of automorphisms
which acts on $I_n$.
In this example the group of automorphisms is $SU(n)$.
In heterotic compactification when $V$ is used to gauge $E_8\times E_8$,
the automorphism group is part of an unbroken gauge symmetry.
This implies that  in
the dual F-theory there is a 7-brane which carries this particular gauge
symmetry group.

Now consider the deformations of $M$ {\it preserving} the multiple
component  $\S'$ of the
spectral surface. Such deformations are described exactly by the
moduli of the bundle $M$ on $\S'$.  In general the deformations break the
structure of the product $E \otimes I_n$ so the automorphism group of
$V$,
which is gauge symmetry in 4 dimensions, can become smaller or disappear
altogether. This should be compared to the symmetry breaking mechanism by
the instanton background inside the 7-brane.

This example shows that the bundle $M$ on the spectral surface is in a
one-to-one
correspondence with the gauge bundle inside the
7-brane in F-theory.
In particular the Chern classes of the two bundles are related. The
precise map between the bundle
on the spectral surface $\S'$ and the gauge fields inside 7-branes
can be quite
complicated. We are planning to return to this discussion in one of our
future publications \fut.
In the section 4 of this paper we will find this map in one simple but
very important example describing the 3-(5-)brane-instanton transition.

We see that the full moduli space of  vector bundles on the  
elliptic Calabi-Yau
threefold is a huge web  which contains various irreducible components.
Moreover, this space has  a natural stratification.
Each stratum is characterized by the number of irreducible components of
the spectral surface, their multiplicities and the
second Chern classes $c_2(M_j)$
of the spectral bundles on multiple components.
The strata are connected through the
transition points.  Some of these transitions are discussed below.

\subsec{Singularities of elliptic fibration}

F-theory is compactified on elliptic Calabi-Yau fourfold with a
section.
 For
practical reasons we represent the elliptic fibration in the
Weierstrass form
\eqn\weis{y^2=x^3+x f(\cdot )+ g(\cdot )~,}
where $f(\cdot)$ and $g(\cdot)$ are the polynomials on the base.
The base of the elliptic fibration is a
complex manifold, which has the structure of $\P^1$ bundle over
$\F_n$.  Let us denote the coordinate along the fiber as $z$,
and the coordinate along the base as $w,u$. The polynomials $f(\cdot)$
and $g(\cdot)$ have the following expansions
\eqn\weexp{f(z,w,u)=\sum_{a=1} ^8 z^a f_a(w,u)~,~~g(z,w,u)=\sum_{b=1}
^{12} z^b g_b(w,u).}
As  suggested in \MV, the polynomials with $a<4,~b<6$ and
$a>4~, b>6$ encode the information about the bundles $V_{1,2}$.
Polynomials $f_4(w,u)$ and $g_6 (w,u)$ govern the complex moduli
of Calabi-Yau threefold of the heterotic compactification.

The singularity of the elliptic fibration along a section
$z=z(w,u)$  can result in the  perturbative gauge symmetry observed in
heterotic
compactification.  It depends on the gauge bundle inside the  
7-brane whether
the full symmetry is observed.
If the bundle is trivial, the singularity structure governs the gauge
group, otherwise the gauge group is broken by instantons.

In the above example
the 7-brane covers the base space $F_n$.
The other possibility for the singularity of the elliptic
fibration is to occur along the divisor $D$ that projects to a  
curve $X$ on the
base. Such singularity corresponds to a nonperturbative effect
on the heterotic side.
Namely, when the discriminant has zero of order greater than $1$
along $D$,  the heterotic string contains a 5-brane wrapped around
the curve $X$ \MV\kucha.

The description of vector bundle on elliptic threefold in terms of  
the spectral
surface $\S$ equipped
with the bundle allows us to identify the degrees of freedom that
correspond to the complex structures on the F-theory fourfold $CY_4$.
Consider the situation  when $E_8 \times
 E_8$ is broken by  bundle $\V= V_1\oplus V_2$ down to  $G_1 \times G_2$.
Similarly to the six-dimensional compactifications, the  
decomposition of the
second Chern class $c_2(\V)$ into $c_2(V_1)$ and $c_2(V_2)$ is   
fixed by the
F-theory data:
\eqn\secch{c_2(V_{1,2})=x_{1,2} AB+(12 +6n \pm m) AS + (12 \pm k) BS~.}
In the case when $n=0$, one can easily derive this decomposition
by restricting the vector bundles $V_{1,2}$ on the cycles $A$ and $B$
both representing $K3$.
Elliptic $CY_3$ in question is a {\it double} $K3$ fibration.
The vector bundles $V_{1,2}$ can be described by fibering the restriction
of the bundle on either of these $K3$s.
Therefore the decomposition of
the Chern classes along both $K3$s implies \secch\ (see \MV).
This arguments can also be generalized for $n \neq 0$ (cf. the  
explanation
after \clsche ).
The only unfixed coefficient is the one in front of $AB$ (projection
along the elliptic fiber).
The sum of these two coefficients is related to the number of
5-branes wrapped
around the elliptic fiber in the compactification in question.
Namely\foot{The second chern class
of the tangent bundle of elliptic Calabi-Yau is equal to
$c_2(T_{CY})=92 A B+(24+12 n) A S+24 B S $.},
\eqn\anomal{N({\rm branes})=c_2(T_{CY_3})_{AB} -x_1-x_2
={ \chi(CY_4) \over 24}- \tilde c_2~,}
where $c_2(T_{CY_3})_{AB}$ is the coefficient in front of $AB$ in the
decomposition of $c_2(T_{CY_3})$ with respect to a basis $AB,~AS$  
and $BS$.
Expression \anomal\ equates the number of 3-branes in F-theory
compactification
with the number of 5-branes in the heterotic compactification.
In the case of singular Calabi-Yau threefold the Euler character can be
computed using the methods discussed in  \klemm. The
last term on
the r.h.s. counts the number of instantons inside 7-branes.

Two bundles $V_{1,2}$ enter on the equal
footing and therefore
we may discuss just one of them.
It follows from \clsche\  that the class of the surface  
$\Sigma(V_1)$ is equal
to
\eqn\Class{\Sigma(V_1)=r_1 S +(12+k)B+(12 + 6n+m) A.
}
{}From the equation \Class\ it follows that  $\S(V_1)$ is a zero  
set of the
polynomial (cf.~eq.~(2.6) in \WM)
\eqn\hyp{
a_0 + a_2 x + a_3 y + a_4x^2 + a_5 x^2y + \cdots,
}
the last term is either $a_rx^{r\over 2}$ for even $r$  or $a_r  
x^{r-3\over 2}
y$ for odd $r$. The coefficients $a_q$ are the sections of the line  
bundle
$K_B^{q-6}\otimes {\cal O}(m\,a+k\,b)$ on the base $F_n$, the line bundle
$K_B={\cal O}(-2b-(n+2)a)$ is the canonical bundle on $F_n$. In  
other words,
the coeffficients $a_q$ are polynomials in $w,\ u$ given by
\eqn\coe{
a_q(w,u)=\sum_i^{12-2q+k} w^i \sum_j^{12-4i+(6-2i)n+m} a_{qij} u^j.
}

The number of complex deformations of the
spectral surface $\S$ is given by the adjunction formula \class.
To illustrate our point let us  compare the number of complex
deformations
in  F-theory with the number of complex
deformation of  the spectral surface $\Sigma$ on the heterotic side.
We summarize the number of complex deformations {\it plus $1$}
in the table below.
These numbers are derived under the assumption that the
spectral surface is
irreducible.

\bigskip
{}~~~~~~~~~~~~~~~Table 1.
\bigskip
\begintable
  | Unbroken subgroup | ${1 \over 12 } (2\S^3+\S \cdot c_2(T_{CY}))$ \elt
$r=1$ | $E_8$ | $169 + 13(k+m) +  k m -  nk (13+  k)/2$ \elt
$r=2$ | $E_7$ | $250 + 22(k+m) + 2 k m -  nk (11+  k)$ \elt
 $r=3$ | $E_6$ | $299 + 29( k+m)+ 3 k m -nk(29+3k)/2$ \elt
 $r=4$ | $SO(10)$ | $ 324 + 34( k + m)+ 4 k m -  nk (17+ 2 k)$ \elt
 $r=5$ | $SU(5)$ | $333 + 37( k + m)+ 5 k m - nk (37 + 5k)/2 $ \elt
$r=6$ | $SU(2)\times SU(3)$ | $334 + 38 (k +  m) + 6 k m - nk(19 - 3 k)$
\elt
$r=7$ | $SU(2) \times U(1)$ | $335 + 37(k + m) + 7km - nk(37 - 7k)/2$
\endtable
\bigskip

Let us compare these expressions with the similar computations on the
F-theory side.
Elliptic fibration is given in Weierstrass form \weis, where
\eqn\weir{\eqalign{f(z,w,u)= \sum_a z^a  \sum_i  ^{8+k(4-a)} w^i  \sum_j
^{8+m(4-a)+n(4-i)}  u^j f_{aij} ~,\cr
g(z,w,u) =  \sum_b z^b  \sum_i  ^{12+k(6-b)} w^i  \sum_j
^{12+m(6-b)+n(6-i)}  u^j    g_{bij}  ~.\cr} }
Below we compute the complex deformations {\it preserving}
a particular singular locus. In doing that we just compute the number of
deformations of polynomials (number of coefficients) that does not affect
the singularity structure.
The number of coefficients differs from the
number of complex deformations by $1$, which is due to the possibility
of rescaling the $z$ coordinate; this does not affect the position
and the structure of the singularity.
For this reason  Table 1 gives the
number of complex deformations {\it plus $1$}.

Note that one can identify the Kodaira type of a singular fiber in
an elliptic fibration by using Tate's algorithm \ref\tate{J. Tate,
{\it Algorithm for determining
the type of a singular fiber in an elliptic
pencil}, in {\it Modular Functions of One Variable IV},
Lecture Notes in Math., vol. 476, Springer-Verlag, Berlin (1975).}.
In the examples considered in this section the conditions for having
a {\it split} singularity turn out to be the same as in 6  
dimensions \kucha .
This determines the unbroken gauge groups as given in Table 1.

We start with the most singular case when elliptic fibration has
$E_8$ singularity located at zero section ($z=0$).
The singularity located at the section at infinity corresponds to
the other bundle, say $V_2$ and is irrelevant for our considerations.
The singularity is characterized by
polynomials  $f_a (w,u)$ and $g_b (w,u)$ with $a \geq 4,~ b \geq 5$.
The number of complex deformations {\it preserving}  $E_8$ locus is equal
to the number of coefficients  in $g_5 (w,u)$ (the rest of the
polynomials specifies other data) and it is given
by
\eqn\eeight{\sum_{i=0} ^{12+k} (13+m+n(6-i))=169 + 13(k+m) +  k m -  nk
(13+  k)/2~,}
provided that $12+m \geq 6n+kn$. Comparing this with \coe\ we see  
that one can
identify $g_5(w,u)$ with $a_0(w,u)$ in \hyp.

As one can see from Table 1, the  $E_8$ singularity in F-theory formally
corresponds to the rank 1  heterotic bundle. To understand this special
situation, let us return to the F-theory/heterotic duality in six  
dimensions.
The $E_8$ singularity on the F-theory side is interpreted there in  
terms of the
zero size $E_8$ instantons on the heterotic side \MV .
The new physics could be described by  tensionless strings. Let us  
see how the
zero size instantons appear in  the spectral theory of $K3$. The  
spectral curve
in the dual $\check{K3}$ for $r=1$ is given by $\S=S+pF,\ p=12\pm  
n$. It is
reducible: one irreducible component is the zero section $S$ and  
$p$ other
irreducible components are fibers $\check{T}^2_i,\ i=1,\ldots ,p$.  
The elliptic
components carry  line bundles $L_i\in T^2_i$ which can be  
identified with
points on the fibers $T^2_i$ of the physical $K3$.   This spectral data
corresponds not to a bundle, but to a torsionless sheaf  with pointlike
instantons (5-branes) located in the points $L_i$ on the fibers
 $T^2_i$. Deformations of the spectral curve $\S$ move the 5-branes  
along the
base \Pt1. Deformations of the spectral bundle $L_i$ move the i-th  
5-brane
along the fiber.  The dimension of the  moduli space ${\cal M}_\S$  
is $p$ which
coincides with the formal genus of $\S$. The full moduli space is  
birational to
the symmetric product ${\rm Sym}^p K3$ and has the dimension $2p$.

Now let us return to four dimensions. We will interpret the $E_8$   
singularity
of $CY_4$ in terms of the instanton (5-brane) wrapped around the curve
$C=(12+6n+m)AS+(12+k)BS$ in the base ${\bf F}_n$. The spectral  
surface $\S$ of
this bundle has two irreducible components. One is the zero section  
$S$ which
is rigid. The other is a preimage of the curve $C$. The  
deformations of $C$
along the base $F_n$ are described by the coefficients in the polynomial
$g_5(z,w)$. Their number is given by \eeight.

For the heterotic bundle of rank $2$, one expects to get the
$E_7$ singularity in the F-theory compactification.
According to Kodaira classification in the case of $E_7$ singularity
all terms $f_a (w,u)$ and $g_b (w,u)$ with $a < 3$ and $b < 5$ vanish.
Again,  we assume that $n,m,k$ satisfies some relations,
namely $8+m \geq 4 n+k n~,~~ 12+m \geq 6 n+k n$.
The deformations are described by the polynomials $g_5(w,u)$ and  
$f_3(w,u)$
which one can identify with $a_0(w,u)$ and $a_2(w,u)$ in \hyp. Thus  
the $SU(2)$
spectral surface corresponding to the $E_7$ singularity is given by
\eqn\sutwo{
g_5(w,u)+f_3(w,u)x=0.
}
In this domain the number of parameters is given by
$$\sum_{i=0} ^{8+k} (9+m+n(4-i))+ \sum_{i=0} ^{12+k} (13+m+n(6-i))$$
which, of course, is the number of deformations for $r=2$
(see Table 1).

For $E_6$ singularity there is an extra term with $g_4 (w,u)$.
This is the first case when the polynomial is not generic and one has to
impose an extra condition (generic polynomial corresponds to
$F_4$ singularity).  The polynomial $g_4(w,u)$ should be a
perfect
square \kucha\  and it could be written as $g_4(w,u)=q(z,w)^2$, where
\eqn\coeff{ q(w,u)= \sum_i  ^{6+k} w^i  \sum_j ^{6+m+n(3-i)}  u^j
q_{ij} ~.}
The polynomial $q(w,u)$ can be identified with $a_3(w,u)$, so the $SU(3)$
spectral surface corresponding to the $E_6$ singularity is given by
\eqn\suthree{
g_5(w,u)+ f_3(w,u)x+ q(w,u)y=0.}
This polynomial $q(w,u)$ produces $\Delta=49+7(k+m)+km-nk(7+k)/2$
 extra parameters,
provided that $6+m \geq 3n+kn$. One can easily see that $\Delta$ is
exactly the
difference between $r=3$ and $r=2$ (see Table 1.).
It is remarkable that the conditions on the polynomials of the elliptic
fibration found in \kucha\ do not depend on the dimension of
compactification.

Let us also check the rank $r=4$ bundle (corresponding to
$SO(10)$ singularity). There are
two additional terms to take into account $f_2 (w,u)$ and $g_3 (w,u)$.
These terms are not independent but should be related
$f_2 (w,u)=h^2(w,u)$ and $g_3 (w,u)=h^3(w,u)$, where
$h(w,u)$ has the following expansion
\eqn\coefso{ h(w,u)= \sum_i  ^{4+k} w^i  \sum_j ^{4+m+n(2-i)}  u^j
h_{ij} ~.}
Sure enough, we can identify $h(w,u)$ with $a_4(w,u)$, so the  
corresponding
$SU(4)$ spectral surface is
\eqn\sufour{
g_5(w,u)+ f_3(w,u)x+ q(w,u)y + h(w,u)x^2=0.}
The polynomial $h(w,u)$ gives $25+5(k+m)+km-nk(5+k)/2$ extra parameters.
Again, this is consistent with the results in Table 1.

Using the results  of \kucha\ one can also verify that the number of
deformations preserving the gauge symmetry enhancement locus
matches with the number of deformations of vector bundles of rank
$r=5,6,7$. These calculations are straightforward and we do not  
present them
here.

On physical grounds, one expects that
the numbers of complex deformations in Table 1 should be
consistent with Higgsing
$E_7\to E_6\to SO(10)\to SU(5)\to SU(2)\times SU(3)\to SU(2)\times U(1),$
similarly to the six-dimensional case \kucha.
However, it is easy to check  that for consistency in four dimensions
one has to assume the existence of (quite nontrivial) superpotentials
in the low-energy effective theory, which by F-flatness conditions would
decrease the dimensions of Higgs branches.
It would be interesting to investigate such superpotentials both in  
heterotic
theory and in F-theory.

In the computations discussed above we assumed that parameters $n,m$ and
$k$
satisfy some conditions. If these conditions are not satisfied then the
summation limits in \weir\ become different.  For simplicity, consider
the case of $n=2$. It turns out that the both conditions
discussed above are equivalent to $m \geq 2k$.
If instead $m < 2k$, then the expansions \weir\ read as follows
\eqn\weirmod{\eqalign{f(z,w,u)= \sum_a z^a  \sum_i  ^{8+2m-[m a/2]} w^i
\sum_j
^{16+m(4-a)-2i}  u^j f_{aij}\cr
g(z,w,u) =  \sum_b z^b  \sum_i  ^{12+3m-[m b/2]} w^i  \sum_j
^{24+m(6-b)-2i}  u^j    g_{bij} ~, \cr~}}
where $[x/2]$ denotes the integer part of $x/2$.  It is clear that the
number of complex deformations  is going to be different from the  
one computed
above under the assumption $m\geq 2k$. We suggest the
following explanation of this phenomenon. When $m < 2k$  the  
spectral surface
becomes reducible. The class of the spectral
surface is given by \Class, but now it has two components $\S'$ and  
$\S''$:
\eqn\comp{\S'=rS+ (24 +m)A +(12+m/2)B ~,~~\S''=(k-m/2)B ~.}
It turns out that the surface $\S''$ is rigid and
therefore all deformations come from $\S'$.   One can easily verify  
that the
number of deformations encoded in polynomials
\weirmod\
exactly matches the number of deformations
of the surface $\S'$.  The example presented here is very simple.
In general, the spectral surface may have several
irreducible components.
It would be interesting to investigate this further.

\subsec{Singularity vs. gauge group}

Below we consider an example in which the 4d gauge group differs  
from the one,
prescribed by the singularity. This example was constructed by
M. Bershadsky, S. Kachru, V. Sadov and C. Vafa (unpublished).
The F-theory is compactified on generalized Hirzebruch with  
\foot{In order to
obtain  a nonsingular heterotic
threefold, one needs to choose $n=0,1,2$.}
$m=12+6n$ and $k=12$.  In this case the elliptic fibration has an
$E_8$ singularity along the section at infinity.  The $E_8$  
component does not
intersect  other components of the
discriminant locus.  We will consider a situation when  that $E_8$ is not
broken by  instantons and the gauge group in four dimensions is  
$SU(5) \times
E_8$.  The factor $SU(5)$  corresponds to the 7-brane {\it with $E_7$
singularity}  wrapped around zero section.  The point here is that  
this 7-brane
carries a nontrivial rank $3$ instanton bundle so that
a would-be $E_7$ is broken to $SU(5)$.

To describe this theory in the heterotic language we note that for
$m=12+6n$  and $k=12$ the decomposition of the
second Chern classes is given by \secch\
\eqn\dec{\eqalign{c_2(V_1)=x_1 AB+(24 &+12n) AS+24 BS \cr
c_2(V_2)&=x_2AB~,\cr}}
where $x_1+x_2=(92- N_5)$, where $N_5$ is the number of 5-branes.
Since we want an unbroken $E_8$ in four dimensions, we choose the  
bundle $V_2$
to be trivial, so $x_2=0$ .  Also we want a {\it perturbative} heterotic
compactification with no 5-branes which forces the condition  
$x_1=92$. Finally,
$c_2(V_1)=c_2(CY_3)$.

The bundle $V_1$ is characterized by assigning to it
the same toric data as
for Calabi-Yau threefold
\eqn\btor{n_{iJ}=\pmatrix{
1 & 1 & n & 0 &4+2n & 6+3n&0  \cr
0 & 0 & 1 & 1 &  4 & 6& 0\cr
0 & 0 & 0 & 0  & 2 & 3& 1\cr
}~,}
where the index $J \in (1,2,3)$.  Let us also define $m_J=\sum_i n_{iJ}$.
The vector bundle is defined by the cohomology of the  sequence
\eqn\exact{0 \rightarrow {\cal O} \rightarrow
\bigoplus_i  {\cal O}(\sum_J  n_{iJ}X_J)
\rightarrow  {\cal O}( \sum_J m_J X_J) \rightarrow
0}
The classes $X_J$ represent the familiar basis $A,B$ and $S$ in
$H^4(CY_3)$. The map from $\bigoplus_i  {\cal O}(\sum_J  n_{iJ}X_J)$ to
${\cal O}( \sum_J m_J X_J)$ is given by the polynomials $F_i(\cdot)$ of
three-degree
$(m_1-n_{i1},m_2-n_{i2},m_3-n_{i3})$.
It is easy to see that $ch(V)=2+ch(CY_3)$ so one can think about
this bundle
as a
deformation of the tangent $SU(3)$ bundle into an
$SU(5)$ bundle.
The gauge group $E_8\times E_8$ is broken down to $SU(5)\times
E_8$.

It turns out that the spectral surface for this bundle consists of
two components $\S'$ and $\S''$, where
\eqn\comp{\S'=2S+(24+12n)A+ 24 B}
and the second component $\S''$ is  a zero section $S$ with
multiplicity $3$. To see that one restricts \exact\ to the elliptic
fiber realized as a degree $6$ hypersurface
$$W_6=y^2-x^3-f\, xz^4-g\,z^6=0$$
in the weighted projective space $W{\bf P}^2_{1,2,3}$. The sequence  
\exact\
becomes
$$0\rightarrow {\cal O}
\rightarrow{\cal O}\oplus {\cal O}\oplus {\cal
O}\oplus {\cal O}\oplus {\cal O}(1)\oplus{\cal O}(2)\oplus{\cal
O}(3)\rightarrow {\cal O}(6)\rightarrow 0,$$
where the second map is given by polynomials $(E_1, ...E_7)$ of degrees
$(0,0,0,0,1,2,3)$ and the map into ${\cal O}(6)$ given by polynomials
$(F_1,\ldots,F_7)$ of degrees $(6,6,6,6,5,4,3)$ such that
$\sum E_i F_i=0~ ({\rm mod}~W_6 )$. It is easy to check that by making
appropriate field redefinitions in the linear $\sigma$-model  
$(F_1,\ldots,F_7)
\rightarrow
({\tilde F}_1,\ldots, {\tilde F}_7)$ one can make ${\tilde F}_1=0,\ldots,
{\tilde F}_4=0$.  That implies that the spectral surface reduces to two
components one of which is the zero section with multiplicity $3$.

It is important that deformations of the spectral
surface are given only by the deformations of the first component
$\S'$, since $\S''$ is rigid.
Therefore, at the level of parameter
counting, this example
coincides with the one of the rank $2$  bundle in spite of the fact
that the rank of the bundle is $5$.
The counting for rank 2 was done in
section 3 where we found that
on the F-theory side this situation corresponds to $E_7$
singularity of Calabi-Yau fourfold!

The component $\S''$ of the spectral surface carries a rank $3$  
spectral bundle
which should be related to the instanton bundle on the $E_7$  
D-brane. This
bundle breaks $E_7$ to $SU(5)$.  The deformations of $V_1$ comes from
1) the deformations of $\S'$ and 2) the deformations of the  
spectral bundle on
$\S''$.  Only the deformations of the first type can be counted using the
Poincare polynomial technique.

\newsec{Mixed moduli space of F-theory in four dimensions}

Here we shall discuss the branches of  the moduli space  of F-theory
compactified on
a fourfold $CY_4$. The possibility for various branches occurs when
$CY_4$ has a
singularity due to  degeneration of elliptic fiber along a
component of the
discriminant locus. The 7-brane corresponding to that component carries
a non-abelian gauge group. The moduli we want to discuss  describe
instantons on
that 7-brane.

The Calabi-Yau manifold $CY_4$ is an  elliptic fibration over the
base $B_F$. If
$CY_4$ is {\it also} a $K3$ fibration over the base $B_H$, it is
conjectured to
be dual to heterotic compactification on $CY_3$ --- Calabi-Yau elliptic
fibration over $B_H$ \ref\vdu{C. Vafa, Nucl. Phys. {\bf B469}
(1996) 403.}
\ref\wdu{E. Witten, {\it Non-perturbative superpotentials in String
Theory}, Nucl. Phys. {\bf B 474} (1996) 343.}. We assume that  
elliptic and $K3$
fibration
structures
are compatible so that the threefold $B_F$ is a \Pt1\ fibration
over $B_H$.

F-theory on $CY_4$ develops anomaly given by $\alpha=\chi/24$ where
$\chi$ is
the Euler character of  $CY_4$ \wvs. To cancel the anomaly  one can put
$\alpha$
3-branes inside $B_F$. In the heterotic theory this means putting
inside $CY_3$
$\alpha$ 5-branes wrapped around elliptic fiber. One expects a one-to-one
correspondence between 3- and 5-branes in two theories.  Therefore
it is very
instructive to compare the moduli space of 3-branes inside $B_F$ with the
moduli of 5-branes inside $CY_3$.

Let us start with smooth $CY_4$ and cancel the anomaly by
$3$-branes. In the
heterotic theory $E_8\times E_8$ is completely broken by the
bundle $\V$.
Choose a complex structure on $CY_4$. Then the moduli space of a
3-brane is
3-dimensional: it is $B_F$ --- the base of the elliptic fibration
$CY_4$. To
see this moduli space in the 5-brane picture we recall that  $B_F$
is a \Pt1\
bundle over $B_H$.  The  {\it position} of a 5-brane is specified
by a point on
a 2-dimensional $B_H$. The coordinate of a 3-brane along the fiber
\Pt1\ should
be
identified\foot{There is a good understanding
of such  identification for the $SO(32)$
heterotic string 5-brane
which carries a $SU(2)$ vector multiplet and a half-multiplet in
${\bf (2,32)}$.
The projective line \Pt1\ can be identified with the moduli space of the
background $SU(2)$ bundle on $T^2$. In four points on \Pt1,
corresponding to
four spin structures on $T^2$, the $SU(2)$ symmetry is restored {\it
classically}. In quantum theory these points correspond to pure
$N=2$  $SU(2)$
gauge theory.  So on the {\it quantum} \Pt1\ each spin structure
gives rise to
 two points separated by the mass gap $\Lambda_i$, $i=1,\ldots,4$.
Also on
\Pt1\ there are 16 points where the Wilson line of $SU(2)$ restores
 a $U(1)$ subgroup  of
$SO(32)$.
A massless charged field which is a part of ${\bf (2,32)}$ appears  
at these
points.
Altogether,
there are $16+4\cdot 2=24$ special points on \Pt1, as expected from
3-5-brane
correspondence. The mass gaps $\Lambda_i$ are not all independent,
because
three nontrivial spin structures of the fiber are permuted by
monodromy around
the discriminant locus. In fact, they fit into a surface which
covers the base
$B_H$ three times. So only one of three $\Lambda_2, \Lambda_3,
\Lambda_4$ is an
independent parameter. Together with $\Lambda_1$ and $16$ Wilson
lines this
gives $18$ independent parameters, also as expected. Tuning $16$
Wilson lines of $SO(32)$
one can find various theories with extended global symmetries.
For example, consider a  $SO(32)$ bundle on $T^2$  given by
$I_8\otimes(L_1\oplus L_2\oplus L_3\oplus L_4)$ where $I_8$ is a trivial
$SO(8)$ bundle and $L_i$ are the four spin structures. The quantum  
moduli space
is \Pt1\ with four special points corresponding to the spin  
structures, where
the global $SO(8)\subset SO(32)$ is restored. At these points, the  
effective
theory is the (finite) $N=2$ $SU(2)$ with four flavors.}
with the position on the Coulomb branch of the moduli space of
5-brane compactified on $T^2$.

3-branes in F-theory are probes measuring the local geometry of elliptic
fibration. Similarly, 5-branes are the probes in heterotic
compactification,
measuring geometry and the restriction $\V \mid_{T^2}$ of the
heterotic  bundle
$V$ to elliptic fiber. Compactified on a given fiber $T^2$ with
given Wilson
lines $\V\mid_{T^2}$, 5-brane has a 1-dimensional Coulomb moduli space
parameterized by the superpartner of photon.

It should be noted that the effective 4-dimensional theory on the
3-brane probe has
$N=2$ supersymmetry for the 3-brane located close to the 7-brane.
The supersymmetry can be broken (by the background) to $N=1$
when a 3-brane approaches special
divisors on 7-branes \ref\aks{O. Aharony, S. Kachru and E. Silverstein,
{\it New N=1 Superconformal Field Theories in Four Dimensions from  
D-brane
Probes},
hep-th/9610205.}. In the
present
discussion we restrict ourselves to generic situation.

Now we have a setup to describe other branches of the moduli space of F
theory on $CY_4$. On these branches, the 3-brane anomaly is  
cancelled by both
3-branes and instantons inside 7-branes. To study the transition from the
no-instantons branch to a branch with instantons let us consider a  
3-brane
probe in the vicinity of the 7-brane carrying a nonabelian gauge  
group $G$. The
effective
theory possesses the product gauge group $U(1) \times G$. Open strings
connecting 3-brane with 7-brane produce a matter hypermultiplet  
charged with
respect to both $U(1)$ and $G$, with mass proportional to the  
distance between
these D-branes. When the 3-brane probe is away from the 7-brane,  
the matter
fields are heavy so that the $N=2$ supersymmetric $U(1)$ theory on  
the probe is
on the Coulomb branch.  As the 3-brane approaches the 7-brane, the
$G-$multiplet of  $U(1)$ hypermultiplets becomes light and a  
transition to the
Higgs branch is possible. Note that the $U(1)$   Higgs branch   
intersects the
Coulomb branch in points where the
number of massless hypermultiplets (the dimension of the  
$G$-multiplet) is  at
least two.  Therefore, the group $G$ has to be at least $SU(2)$.

To make the $G$-multiplet of matter fields massless, the 3-brane  
should sit on
top of the 7-brane. It is known that such configuration of D-branes  
can be
identified with a
point-like $G$-instanton. Turning on
the $vev$ of the matter hypermultiplets means
smoothing out
the singular gauge fields
corresponding to the point-like instanton \ww \doug.
Therefore, the Higgs
branch of the effective theory consists
of instantons of finite size of the nonabelian gauge group $G$
on the 7-brane. At the transition point the $U(1)$ gauge group
decouples (at least for $G=SU(n)$) and
the Higgs branch is in fact the Higgs branch of the gauge theory  
with the gauge
group $G$.

To sum up, the transition amounts to eating up $k$ 3-branes and
turning on $k$ instantons inside
the 7-brane with nonabelian gauge group $G$. The F-theory anomaly remains
cancelled.

Consider an important example when the compact part of the 7-brane  
worldvolume
 is the zero section of the bundle
$B_F\rightarrow B_H$, so  it can be identified with $B_H$ itself.
Suppose that
the singularity of the elliptic fibration along this 7-brane is such
that in the
absence of instantons there is a gauge  group $G$ in four  
dimensions.  Let us
take  $k$ 3-branes on top of this 7-brane, so that the ``3-brane  
group'' is
$SU(k)$.
Each 3-brane produces a hypermultiplet
(a pair of chiral fields).
These hypermultiplets are massless states of
the open strings connecting 3-branes and 7-branes. The ``3-brane end''
of the string carries the flavor index of $SU(k)$, while the  
``7-brane end''
is charged with respect to the gauge group on the 7-brane.
Giving $vev$ to the hypermultiplets, one makes the 3-brane-instanton
transition. In the 4-dimensional field theory language, the gauge  
group gets
broken
 by the nonzero $vev$'s.   In the D-brane language, it is broken by the
 $G$-instantons on the 7-brane.

These two descriptions
should match.
For $G=SU(n)$
the dimension of the instanton moduli space
is equal to $2 n k-(n^2-1)\,(1-h^{01}+h^{02})$.
Taking into account that for $B_H$ the Hodge numbers  
$h^{01}=h^{02}=0$, we
arrive at
\eqn\higdim{{\rm dim}{\cal M}=2n k-(n^2-1) ~.}
This formula has indeed a very clear interpretation in
terms of Higgs mechanism \bjsv : it counts the dimension of the
Higgs branch of $SU(n)$ gauge theory\foot{For a recent discussion
of $SO$ and $Sp$ cases
see ref. \zw .
A consideration of exceptional groups would involve tensionless strings
\MV.}.
The instanton number $k$ coincides with the number of
3-branes
and therefore counts flavors: $N_f=k$ since each 3-brane produces
two chiral
fields ($N_f$ counts the number of pairs of fundamental-antifundamental
representations).

When $N_f<n$ there are no flat directions
due to the nonperturbative
superpotential  $\sim1/(Q\tilde{Q})^{1/(n-N_f)}$
where $Q$, $\tilde{Q}$ are squark
chiral superfields.
The appearance of such
superpotentials is well known
in field theory
\ref\superpot{I. Affleck, M. Dine and
N. Seiberg, Nucl. Phys. {\bf B241}  (1984) 493;
Nucl. Phys. {\bf B256}  (1985) 557.}.
In the case of massless squarks this superpotential lifts the  
ground state of
the
field theory.
When squarks have finite non-zero masses the theory has a stable vacuum
corresponding to nonzero expectation values of squarks.
In the context of F-theory for $N_f=n-1$,
this superpotential was recently discussed in \bjsv .
The bare masses of squarks are proportional to distances
between 7- and 3-branes.
Therefore at the level of effective field theory the limit of vanishing
distance between 7- and 3-branes is ill-defined and may require
various quantum corrections that stabilize\foot{Note that the field  
theory
on the 3-brane being IR free does not
stabilize the ground state.
The question of back reaction from 7-branes to 3-brane probes
may be relevant in this context.} the vacuum
of F-theory at
$<Q\tilde{Q}>\sim M_s^2$, where $M_s$ is the string scale.
In the context of $(0,2)$ heterotic compactifications,
this question was recently discussed in
ref.~\ref\maybe{S. Kachru, N. Seiberg and E. Silverstein,
{\it SUSY Gauge Dynamics and Singularities of 4d N=1 String Vacua},
Nucl. Phys.  {\bf B480}  (1996) 170.}.

In general, apart from the non-chiral matter coming
from 3-branes (we call this matter
Type A), there are  chiral matter fields (we call them Type B) which
come from
the intersections of the given 7-brane with other 7-branes\foot{
Clearly, we are trying to be simplistic here.
The intersection of 7-branes may require appropriate resolutions
(or blowups), similar to the ones discussed \ref\coll{M. Bershadsky and
A. Johansen, {\it Colliding Singularities in F-theory and Phase
Transitions}, hep-th/9610111.}.  Already in six dimensions some  
collisions
did not have an interpretation in terms of conventional field theory  and
require tensionless strings. We expect  similar phenomena in
four-dimensional compactifications \ref\vkac{C. Vafa, private
communication.}.}.
In the effective
$N=1$ supersymmetric  4-dimensional theory there is a tree level
superpotential $W=W(A,B)$
which couples these two sorts of matter.
This superpotential implies that when the Type A fields develop
nonzero $vev$'s,
some of the Type B fields may become massive. The origin of this
superpotential can
be  more clearly seen in the D-brane language. Nonzero $vev$'s  of
the Type A
fields correspond to the nontrivial instanton background in the compact
directions. Type B multiplets are charged with respect to the gauge
group so
that they interact with the instanton field and become massive.

The nonzero $vev$'s of Type B fields also break the gauge symmetry.
In the
D-brane language these $vev$'s generate splitting of the 7-brane
with charge
$Q_{(7)}>1$ to several 7-branes with smaller charges. Concretely,
Type B matter
is a source term in the generalized Hitchin equations describing  
fields in
the bulk of
the 7-brane. We call fields in the bulk Type C matter.
Nontrivial solutions
for these Type C fields which transform in the adjoint of the gauge group
correspond to the splitting of the 7-brane. We will give a more detailed
exposition on this in the section 5.

As an immediate consequence of the above discussion we see that when
the $vev$'s
 of the Type A fields
are turned on (=there is nontrivial instanton background), the
moduli of complex deformations of $CY_4$ which split the
corresponding
7-brane with $Q_{(7)}>1$ are frozen. More precisely, only the
splittings
incompatible with the
instanton embedding are forbidden. For example, a multiplicity two
7-brane with a $SU(2)$ instanton cannot split into two $U(1)$ 7-branes.
Therefore on this branch  of the moduli space the fourfold $CY_4$
is always
singular.

Now let us return to the heterotic picture.
When a 3-brane disappears from F-theory, a 5-brane should disappear from
heterotic theory. This means that in that particular point on the
Coulomb branch
the 5-brane can be interpreted as
a singular gauge field configuration such that the curvature is
zero everywhere
except on one fiber where it has a singularity. If $(z,w)$ is a
pair of local
coordinates on the base so that $z=w=0$ are the equations
describing the fiber,
 the Pontryagin 4-form of the gauge field is ${\rm Tr}F\wedge
F \sim k \delta^{(4)}(z,w,\bar{z}, \bar{w})$.
The Higgs
branch corresponds to smoothing out  this singular configuration
which changes
the bundle $\V$ to a new bundle $\W$. In particular,
$\int_{B_H}c_2(\W)=\int_{B_H}c_2(\V)+k$ so that
$\W$ can take care of anomaly imbalance left when the 5-brane is
removed. The
way the new bundle $\W$  breaks the gauge symmetry  should be the
counterpart of the gauge symmetry breaking by the instanton field on the
F-theory side.

Throughout this discussion we are making an assumption that the
positions of
3-branes are independent from the positions of 7-branes inside $B_F$.
This is essentially a version of the ``probe argument'' of \bds. If
this is
true, the 3-brane-instanton transition should
not change
the distribution of 7-branes, i.~e.~the complex structure of
$CY_4$. It follows
from the discussion in section 3 on F-theory -- heterotic correspondence
 that the spectral curve $\Sigma$ of the heterotic bundle $\V$ is
preserved
in the 5-brane-instanton transition in the dual picture:
$\Sigma(\W)=\Sigma(\V)$.

The only piece of spectral data that is left to change in the
transition is the spectral  bundle $M$, which lives on the
component
$\S$ of the spectral surface with ${\rm mult}(\S)=n>1$. It is clear  
that in
general  $M$ has
moduli (for sufficiently large $c_2(M)$, it does).
This supports
the idea
that the multiple components of
the spectral surface are in a one-to-one correspondence with the 7-branes
carrying
nonabelian gauge groups.
To establish this correspondence at least in one case let us
re-examine in the
heterotic language the example described above in the F-theory language.
Namely, we consider a 3-brane-instanton transition on the 7-brane
which is
wrapped around the zero  section of $B_F\rightarrow B_H$ and carries the
gauge group $G$. We can identify the 7-brane locus with the base  
$B_H$ of the
heterotic fibration $CY_3\rightarrow B_H$.

 The corresponding spectral surface on the heterotic side has several
components. The nontrivial part of the $E_8\times E_8$-bundle
$\V$ is coded
by the surface $\Sigma (\V)$ and the line bundle $L(\V)$. The
unbroken gauge
symmetry $G$
corresponds to the component of the spectral surface which is the
zero section $S$
of the fibration $CY_3\rightarrow B_H$.  Again, the surface $S$ can be
identified with the base $B_H$. This component of the spectral  
surface carries
a trivial $G$-bundle.

We start on the Coulomb branch so there are no instantons on the
7-brane and all
the anomaly is cancelled by 3-branes (5-branes). Now let a 3-brane
approach the 7-brane wrapped around the zero section. As we have  
discussed
above, the Higgs
branch develops and
the 3-brane
dissolves into an instanton on $B_H$. Let us denote the corresponding
background gauge bundle by $\tilde M$. In the heterotic picture, a  
5-brane
develops the
Higgs branch and dissolves into an instanton so that the new
heterotic bundle
is $\W$. The bundle $\W$ has the same spectral surface $\S(\V)$ as the
pre-transition
bundle $\V$. However, on the zero section  component of $\S(\V)$,
a nontrivial spectral bundle $M$ develops.
The example
discussed in section 3 shows that\foot{More generally, $\W$ can be a
deformation
called a Hecke transform of $\V\oplus \pi^* M$ but for now the
difference is not important.}
actually $\W=\V\oplus \pi^* M$. The second Chern class $c_2(M)=k$
counts the 5-branes
dissolved in the transition while $c_2(\tilde M)$ counts the
corresponding 3-branes.
These numbers should be equal. Both $M$ and $\tilde M$ were defined  
as bundles
over $B_H$. Now we can describe the rest of the F
theory-heterotic map: we suggest that simply $M=\tilde M$!

\newsec{Matter in heterotic and F theories}
\subsec{Heterotic string after the 5-brane-instanton transition}

In the previous section we discovered how the heterotic bundle $\V$
changes
when several 5-branes dissolve to become instantons.
The spectral surface of the resulting bundle
$\W$ has a special form $\S(\W)=\S(\V)+nS$, i.e. it is a union of
the spectral surface of $\V$ and the zero section $nS$.
Multiplicity $n$ of the
zero section means e.g. that this component of the spectral
surface carries a rank $n$
bundle $M$. The zero section $S$ is isomorphic to a base of elliptic
fibration and therefore one
can think about $M$ as a vector bundle on a base $B_H$.
Note that we have identified this bundle with
the instanton bundle on the 7-brane.
On the other hand the heterotic bundle $\W$ is (a
deformation of) $\pi^*M\oplus \V$. The moduli of $\W$ are a part of the
massless spectrum of the theory and thus are worth investigating.

To examine the deformations of a bundle of the form ${\cal W} =
{\cal V} \oplus \pi^{*}M$ we look at the space
\eqn\decomp{\eqalign{
H^{1}(End({\cal W})) =&
H^{1}(End({\cal V}))\oplus H^{1}(End(\pi^{*}M)) \oplus \cr
&H^{1}(Hom({\cal V},\pi^{*}M))\oplus H^{1}(Hom(\pi^{*}M, {\cal V})). \cr
}}
 The elements of
$H^{1}(End({\cal V}))\oplus H^{1}(End(\pi^{*}M))$ correspond to
the deformations of ${\cal W}$ that preserve the direct sum
decomposition structure
and deform the two direct summands ${\cal V}$ and $\pi^{*}M$
independently.
The elements of
$H^{1}(Hom({\cal V},\pi^{*}M))$ give
the deformations of ${\cal W}$ that are no longer direct sums but
rather fit
in an exact sequence. More precisely an element $\mu \in H^{1}(Hom({\cal
V},\pi^{*}M))$ gives us an extension
\eqn\exmu{
0 \rightarrow \pi^{*}M \rightarrow {\cal W}_{\mu} \rightarrow {\cal V}
\rightarrow 0
}
which is a deformation of ${\cal W}$. Similarly an element
$\nu \in H^{1}(Hom(\pi^{*}M, {\cal V}))$ corresponds to an
extension
\eqn\exnu{
0 \rightarrow {\cal V} \rightarrow {\cal W}^{\nu} \rightarrow \pi^{*}M
\rightarrow 0
}
which is another deformation of ${\cal W}$.

The decomposition \decomp\ is true universally. On the {\it elliptic}
$CY_3$ we can
understand all three types of deformations in terms of the spectral
data. That
will allow us to find the counterparts of these moduli in F-theory.
Let us
start  the discussion with the deformations of $\pi^*M$.

The deformations of $\pi^{*}M$ are given by $H^{1}(CY_{3}, \pi^{*}
End(M))$. Assume that $M$ is a good instanton bundle on $B_H$ so that
$H^2(B_H,
End(M))=0$. Then one can prove\foot{ By the Lerray spectral
sequence applied to
the fibration
$\pi : CY_{3} \rightarrow B_H$ we have
$$
\dim H^{1}(CY_{3}, \pi^{*}End(M)) = \dim H^{1}(B_H, \pi_{*}\pi^{*}End(M))
+ \dim  H^{0}(B_H,
R^{1}\pi_{*}\pi^{*}End(M))
$$
Now $\pi_{*}\pi^{*}End(M) = End(M)\otimes \pi_{*}{\cal O}_{CY_{3}}$ and
$R^{1}\pi_{*}\pi^{*}End(M) =  End(M)\otimes R^{1}\pi_{*}{\cal
O}_{CY_{3}}$.
Since the fibers of $\pi$ are connected and compact we have
$\pi_{*}{\cal O}_{CY_{3}} = {\cal O}_{B_H}$ and also relative  
duality gives
$R^{1}\pi_{*}{\cal O}_{CY_{3}} = (\pi_{*}K_{CY_{3}/B_H})^{*} =
(\pi_{*}(K_{CY_{3}}\otimes \pi^{*}K_{B_H}^{-1}))^{*} =
(\pi_{*}(\pi^{*}K_{B_H}^{-1}))^{*} = (K_{B_H}^{-1}
\otimes \pi_{*}{\cal O}_{CY_{3}})^{*} = K_{B_H}$. In other words  
there are
two types of deformations of $\pi^{*}M$ parameterized by the spaces
$H^{1}(B_H, End(M))$ (deformations coming from the base) and
$H^{0}(B_H, End(M)\otimes K_{B_H})$ (deformations that are non-trivial
along the fibers).
Since $End(M)$ is isomorphic to its own dual the space
$H^{0}(B_H, End(M)\otimes K_{B_H})$ is dual to $H^{2}(B_H, End(M))$ and
we conclude
that all the deformations $\alpha \in H^1(CY_3,End(\pi^*M))$ of
$\pi^{*}M$ come
from the base if and only if
$H^{2}(B_H, End(M)) = 0$.} that all the deformations of the pullback
$\pi^*M$ on
$CY_3$ come from the deformations of $M$ on the base $B_H$. This
observation
allows us to identify these moduli with the moduli $H^1(B_H,  
End(M))$ of the
instanton background in F-theory.
The number of such moduli is given by the index formula
\eqn\mdefs{-\chi(B_H, M)=2nk-(n^2-1)}
where $k$ is the instanton number of $M$.
We have already discussed this
formula in the F-theory context.

Next, consider the deformations of the bundle $\V$ which is the heterotic
bundle before the transition.
The spectral data for $\V$ consist of the surface $\S(\V)$ and the
line bundle
$L(\V)$. We assume that $\S(\V)$ has no multiple components so that any
deformation $\beta\in H^1(CY_3, End(\V))$ is a deformation of the
spectral surface $\S(\V)$.
As discussed in section 3 in F-theory these deformations
correspond to the complex structures of the Calabi-Yau fourfold $CY_4$.

Finally, the spectral surface of ${\cal W}$ has two components
$\Sigma({\cal
V})$ and $n S$ where $n = {\rm rank} \, (M)$.
Assume that $M$ is well behaved
(i.e., that $H^{2}(B_H, End(M)) = 0$) and let $\widetilde{{\cal W}} =
{\cal V}^{\beta} \oplus \pi^{*}M^{\alpha}$ be a deformation of ${\cal W}$
having an infinitesimal $(\alpha, \beta) \in
H^{1}(CY_3,End({\cal V}))\oplus H^{1}(B_H, End(M))$. From the
discussion above we see that
the spectral surface of $\widetilde{{\cal
W}}$  again has the form $\Sigma({\cal V}^{\beta}) + n S$. Thus,
the bundle
$M^{\alpha}$ and the line bundle $L({\cal V}^{\beta})$ describe
$\widetilde{{\cal W}}$ completely.

If, on the other hand, we have a deformation of the type, say,
${\cal W}_{\mu}$
with $\mu \in  H^{1}(Hom({\cal V},\pi^{*}M))$, then the spectral
surface of ${\cal W}_{\mu}$ is exactly the same as the spectral
surface of ${\cal W}$.
This is obvious since the exact sequence \exmu\ guarantees that
${\cal W}$ and ${\cal W}_{\mu}$ have the same Harder-Narasimhan
filtration when restricted on the general fiber of $\pi$.
So what do the moduli
$H^{1}(Hom({\cal V},\pi^{*}M))$ and $H^{1}(Hom(\pi^{*}M,\V))$ do with the
spectral data?

To explain what is happening recall that a bundle ${\cal W}$ on
$CY_{3}$ is encoded in a pair $(\Sigma({\cal W}), L({\cal W}))$ where
$\Sigma({\cal W}) \subset \check{CY}_{3}$ is a divisor (not necessarily
reduced) and $L({\cal W})$ is a line bundle on $CY_{3}\times_{B_H}
\Sigma({\cal W})$. The bundle ${\cal W}$ is the push-forward of the sheaf
$L({\cal W})$ under the natural projection $CY_{3}\times_{B_H}
\Sigma({\cal W})
\rightarrow CY_{3}$. One important feature of this description is that
by definition the points of $\Sigma({\cal W})$ over a point $b \in  
B_H$ are
precisely the degree zero line bundles that participated in the
associated
graded of ${\cal W}_{\mid\pi^{-1}(b)}$ with respect to its
Harder-Narasimhan
filtration.

If it happens that $\Sigma({\cal W})$ has two
irreducible components $\Sigma({\cal W}) = n' \Sigma' + n''
\Sigma''$, then
${\cal W}$ comes furnished with extra structure.
The restrictions $L'$
and $L''$ of $L({\cal W})$ on $n'\Sigma'$ and $n''\Sigma''$ respectively
correspond say to vector bundles ${\cal V}'$ and
${\cal V}''$
of ranks $n'd'$ and $n''d''$, respectively.
Here $d'$ and $d''$ are the
degrees of $\Sigma'$ and $\Sigma''$ over $B_H$. The bundles ${\cal
V}'\oplus
{\cal V}''$ and ${\cal V}$ coincide outside the divisor $D :=  \pi^{-1}(
\pi (\Sigma' \cap \Sigma''))$. More precisely ${\cal V}$ is a
modification
called a Hecke transform of ${\cal V}'\oplus {\cal V}''$ along $D$.
Specifically
this means that there is an exact sequence of sheaves on $CY_{3}$
\eqn\hecke{
0 \rightarrow {\cal V} \rightarrow {\cal V}'\oplus {\cal V}'' \rightarrow
{\cal Q} \rightarrow 0
}
where ${\cal Q}$ is the vector bundle on $D$ obtained as a
push-forward of
the restriction of $L'$ (or $L''$) on the intersection of the two
divisors
$CY_{3}\times_{B_H} (n'\Sigma')$ and $CY_{3}\times_{B_H}  
(n''\Sigma'')$ in
$CY_{3}\times_{B_H} \check{CY}_{3}$. The exact sequence \hecke\ encodes
the
condition that $L'$ and $L''$ come from a global line bundle on
$\Sigma({\cal V})$. There is one limit case when ${\cal V}$ itself
becomes a
direct sum. This happens precisely when ${\cal Q} = {\cal V}'_{\mid D} =
{\cal V}''_{\mid D}$ and the map ${\cal V}'\oplus {\cal V}'' \rightarrow
{\cal Q} $ is just the difference. In this case we have ${\cal V} =
{\cal V}'\oplus {\cal V}''(-D)$.

Let us now return to the special case $\W=\V \oplus \pi^*M$.
The
deformations of the type ${\cal W}_{\mu}$ \exmu\ and $\W^\nu$ \exnu\
with $\mu \in  H^{1}(Hom({\cal V},\pi^{*}M))$
 and  $\nu \in  H^{1}(Hom(\pi^{*}M,\V))$ do not deform the spectral
surface
$\S({\cal W})$. In fact these moduli can be identified with the
moduli of all
non-trivial Hecke transforms \hecke\  we can perform on
$\pi^{*}M\oplus {\cal
V}(D)$ in order to obtain ${\cal W}_\mu$.

Certain information about the number of such moduli can be obtained
by the
index formula. If we denote $N={\rm dim}H^1(Hom(\V,\pi^*M))$ and
$\bar{N}={\rm
dim}H^1(Hom(\pi^*M,\V)$ then from  duality and Riemann-Roch it
follows that
\eqn\deform{
N-\bar{N}={1\over 2}{\rm rank}(\V)\ c_3(\pi^*M)-{1\over 2}n\  
c_3(\V)=-{1\over
2}n\ c_3(\V).}

Thus for $c_3(\V)\neq 0$ this matter is {\it chiral.} To find these  
moduli  in
F-theory we need to recall that matter can
come from
the intersections of D-branes. For instance, above we have
interpreted massless
multiplets coming from intersections of the 7-brane  with the dissolved
3-branes as the moduli $H^1(B_H,M)$ of the instanton background. We
would like to
relate
the moduli $H^{1}(Hom({\cal V},\pi^{*}M))$ and
$H^{1}(Hom(\pi^{*}M,\V))$ to
multiplets coming from the intersections of the 7-brane wrapped on
$B_H$ with
other 7-branes. To do that we would need to understand in F-theory  
the r\^ole
of Hecke transform
similar to \hecke.

\subsec{Geometry of vector bundles on intersecting D branes}

Let us return to the discussion started in section 4 about the
matter fields in
F-theory. We can classify them by the dimension of their support on
the compact
part of the 7-brane. Fields living in the bulk we call Type C. The
ones living
along curves on 7-branes we call Type B. In the simplest case these
special
curves are simply the intersections of two 7-branes. Finally, the
fields coming
from the points on the 7-branes are called Type A. The only example
of Type A
fields we will consider here are hypermultiplets which come from
intersections with
3-branes.

We need to describe the vacua of the ABC system. The appropriate
setup for
this is given by the generalized Hitchin equations with
the source terms. Without the source terms these equations describe the
Type C fields in bulk.
The source terms introduce
couplings to Type A fields localized in points and to Type B fields
localized along special curves within the 7-brane. To be concrete,  
we discuss
 7-branes with $SU$ gauge groups.

We start with the fields in the bulk. In 8 dimensions there is a vector
multiplet. It has a complex scalar $\Omega$ in the adjoint.
The nonzero {\it vev} of this field  signals the splitting of the  
7-brane  to
several parallel components.
The eigenvalues of $\Omega$ measure the spatial separations of these
components.
After compactification  of
four dimensions on the complex surface $X$ the scalar is twisted so it
becomes a section of the canonical line bundle $K_X$ or just a
$(2,0)$-form  $\Omega$.
The components  of the eight-dimensional connection $A_M$ along
$X$ determine the gauge background.
Without coupling to Type A and Type B fields, the
linearized equations of motion simply tell us that the background  
gauge bundle
$M$ is holomorphic and that $\Omega$ is a holomorphic section of  
$End(M)\otimes
K_X$.
That is, $\Omega \in H^2(X, End(M))$.
If $M$ is a well-behaved instanton bundle,
the space $H^2(X, End(M))$ is empty,
so necessarily $\Omega=0$ which guarantees
that the 7-brane carrying such an instanton bundle cannot split.
This condition should be compared with the condition on the
spectral bundle $\tilde M$ discussed in section 5.1 in the
heterotic picture, which ensured that the multiple component
of the spectral surface is preserved by the deformations of the  
bundle $\W$.

Now let us couple this system to Type A matter coming from a  
3-branes with
charge $Q_{(3)}=k$ sitting on top\foot{We assume that the 3-brane is away
from the special divisors on the 7-brane as in \aks.} of the  
7-brane. Such
3-brane carries a IR trivial $N=2$ theory with the gauge group  
$SU(k)$ and
three multiplets in the adjoint.   Its intersection with the 7-brane
carrying $SU(N)$ gauge group produces  two chiral multiplets $Q$ and
$\tilde{Q}$ in conjugate representations $({\bf k,N})$ and $(\bar{\bf
k,}\bar{\bf N})$ of $SU(k)\times SU(N)$. With respect to
the $SU(N)$ gauge theory on the 7-brane the (decoupled) gauge group
$SU(k)$ is a flavor symmetry.

To get a slightly different angle on the problem,
 let us start with the trivial gauge background $M={\rm id}.$
Trivializing $M$ we can think of the connection $A_M$ just as
of the holomorphic 1-form with values in the adjoint of $SU(N)$.
Also, to the linear  approximation, $\Omega$ is a holomorphic 2-form.
So in the  space-time there will be $h^{0,1}+h^{0,2}$
fields
in the adjoint ($h^{0,1}$ fields $h_a$ and $h^{0,2}$ fields $\Omega_i$).

To get to the next approximation we notice that there is a topological
coupling
of two $(0,1)$ forms with one $(2,0)$ form
\foot{This coupling comes out from
8-dimensional coupling $\Omega\lambda\lambda$ under the dimensional  
reduction,
where $\lambda$ stands for
twisted gaugino.}.
In the effective $N=1$ space-time
theory that gives a superpotential $W=C_{iab}{\rm Tr}(\Omega_i[h_a,h_b])$
with the coefficient $C_{iab}$ determined by the intersection numbers in
cohomology
of  $X$.
The
superpotential $W$ also includes a term $\sum_i B^i \tilde{Q}\Omega_i
Q$ where $B_i$ are constants which couple Type A multiplets $Q$ and  
$\tilde Q$
to $\Omega$.

A theory with this superpotential can be easily analyzed.
The classical vacua are
determined by this superpotential together with the $D$-flatness  
condition
and gauge invariance.  The superpotential $W$ is linear in $\Omega_i$ and
bilinear in $h_a$ and $Q,\ \tilde{Q}$. So the classical moduli  
space has a
branch with
\eqn\moqu{\Omega_i=0,  ~~~B^iQ\tilde{Q}+ \sum_{ab}C^{ab}_i
[h_a, h_b]=0, ~~~i=1,\ldots, h^{0,2},}
which has complex dimension
${2Nk-(N^2-1)(1-h^{0,1}+h^{0,2})}.$
In other words, this space is obtained starting from the complex vector
space of dimension $2Nk+(N^2-1)h^{0,1}$ with coordinates given by the
components of $Q$, $\tilde{Q}$ and $h_a$. In this space we take a  
complete
intersection of $(N^2-1)h^{0,2}$ quadrics and finally divide by the  
action
of the complexified gauge group $SL(N,{\bf C})$.

This space has the same dimension as  the moduli space of $SU(N)$
instantons on $X$ with the instanton number $k$. If $h^{0,2}=0$ the  
moduli
space
of the Higgs branch is
a symplectic quotient with respect to the action of $SU(N)$.  When
$h^{0,2}=1$ (so that $X=T^4$ or $X=K3$) the construction above is a
hyperk\"ahler quotient. For $h^{0,2}>1$ it is a generalization of the
hyperk\"ahler quotient. We conjecture that it gives a local  
description of the
instanton moduli space of $X$ in the vicinity of the point-like
$k$-instanton configuration. Essentially this is a generalization of the
ADHM construction on $T^4$ discussed in the similar context in \ww .

Both Type A and Type C matter fields are non-chiral. Thus the only  
source of
possible non-chirality is Type B matter which comes from   
intersections of
pairs of 7-branes. Type B interacts with Type A and Type C via  
superpotentials
which were qualitatively analyzed in section 4. For example, the  
superpotential
$W(C,B)$ makes sure that the nonzero expectation values of Type B get
translated into a nonzero expectation value of $\Omega$. In turn,  
$\Omega \neq
0$ splits the 7-brane thus providing a D-brane realization of the Higgs
mechanism.

More concretely, suppose that $C\subset X$ is the  curve
where $X$ intersects a 7-brane $Y$ with $SU(K)$ gauge group.
Type B matter multiplets $q$ and $\tilde q$ transform respectively  
in $({\bf
N},{\bf K})$ and $(\bar{\bf N},\bar{\bf K})$ of $SU(N)\times SU(K)$.
Assuming $C$ is smooth, they can be computed as cohomology
groups $H^0$ and $H^1$ of the bundle $M_X\otimes M_Y\otimes {\cal L}$
where $\cal L$ is a twisting line bundle on $C$.
The chirality is measured by the Euler character of this bundle  
which is equal
to $NK(1-g(C)+{\rm deg}{\cal L})$.
The latter formula should be compared, for $K=1$ and $n=N$,
with \deform\ in section 5.1.
 The 2-form $\Omega$ and the gauge connection $A_M$ satisfy
a set of equations  with a $\delta$-functional source term along $C$.
These equations  show in particular that $\Omega$ develops a pole
along $C$ with the residue $\tilde{q}q$
bilinear in the Type B matter fields. In a different context, the  
generalized
Hitchin systems on curves instead of surfaces were also considered by
N.~Nekrasov in \ref\nekr{N.~Nekrasov, {\it Holomorphic Bundles and
Many-Body Systems}, Comm.~Math.~Phys.~{\bf 180} (1996) 587.}.

\newsec{Discussion}

Let us first summarize what we have already learned.
The total moduli space of F-theory compactifications is a  
stratified space
with numerous components.
Each component is characterized by the brane
configuration and the gauge bundles inside 7-branes.
The components are connected through the transition points.
The moduli space of the heterotic compactifications on Calabi-Yau
threefolds is also stratified.  Each component of the moduli space of the
bundles on Calabi-Yau threefold is characterized by the number of  
components
of the spectral surface and their multiplicities. The duality hypothesis
implies that one can identify the corresponding strata of the F-theory
moduli space with the corresponding strata of the heterotic
compactification. The details of this  identification  depend on the
strata.

The generic strata correspond to  F-theory compactifications
on {\it nonsingular}  elliptic $CY_4$  (that means some inequalities for
$(n,m,k)$ that characterize the base of F-theory compactification).
Then the elliptic fibration generically has only $I_1$ singularities.
The gauge group is completely broken
(except for some $U(1)$'s).
Such compactifications are characterized by three integers $(n,m,k)$
\WM.
In the heterotic dual one has to specify two $E_8$ bundles
characterized
by the second Chern classes $c_{2}(V_1)$ and $c_2(V_2)$.
The third Chern classes $c_3$
are identically equal to zero for  $E_8$ bundles.
The sum $c_2(V_1)+c_2(V_2)$  of the
second Chern classes is fixed by the 5-brane anomaly cancellation  
condition,
which leaves three independent integer-valued parameters, also.

When the elliptic fibration has higher singular loci
(with singularity higher than
$I_1$) the situation is different.
We would like to propose the following picture:
First let us consider the case
when the discriminant of elliptic fibration has only one irreducible
component (the 7-brane) with the higher singularity located on
the zero section of $B_F\rightarrow B_H$ (see section 4).
In order to specify the  F-theory compactification,
one has to fix
the gauge bundle $\tilde M$ inside the 7-brane.
This introduces two new integer-valued parameters --- the
rank $r$ of the bundle and the  second Chern class  $c_2(\tilde M)$.
The total number of parameters that characterize this
compactification is equal to $5$.
On the   heterotic side one $E_8$  is completely broken by  a generic
$E_8$ bundle  while the other $E_8$ is broken by a vector bundle $V$.
In general,
the bundle $V$ has the {\it nonzero} third Chern class $c_3(V)$.
The rank and the structure group of the bundle $V$ is fixed by the  
singularity
type on  F-theory side.
The spectral surface $\S(V)$ for  this
bundle is reducible and has a component { with multiplicity} $r$.
This component is equipped with a vector bundle $M$ of rank $r$
with the second Chern class equal to $c_2(M)=c_2(\tilde M)$.
The third  Chern class of
the bundle $V$  is related to  $c_2(\tilde M)$.
Now, the independent parameters on the heterotic side are the three
components  of the second Chern class $c_2(V)$,
 the third Chern class $c_3(V)$
and the multiplicity $r$ of one of the components of the spectral
surface.

In general, both $E_8$ groups  are not broken completely, so that  
that the
residual gauge symmetry group is $G_1\times G_2$. In the corresponding
F-theory compactification the discriminant has (at least) two components
(7-branes) with higher singularity located, say at the zero section and
at the section at infinity of the bundle $B_F\rightarrow B_H$ (see  
section 4).
The full set of data includes also the information about the  
background gauge
bundles $\tilde M_i$ inside these two 7-branes --- their ranks $r_i$ and
their second Chern classes $c_2 (\tilde M_i)$.
Again, the ranks of the bundles $\tilde M_i$ determine the
multiplicities of various components of the spectral surface  while
the Chern classes $c_{2}(\tilde M_i)$ are related to the  Chern  
classes of
bundles $V_i$.
The  parameters that characterize the F-theory
compactification are $n,m,k$, the Chern classes $c_2(\tilde M_i)$ and
the ranks $r_i$ of the bundles inside the 7-branes.
These parameters match with the corresponding parameters on the
heterotic side --- the components of the second Chern classes $c_2  
(V_i)$,
the third Chern classes $c_3(V_i)$ and the multiplicities of   
components of the
spectral surface.

We want to emphasize the striking analogy
between the D-branes in the F-theory description and the spectral  
surfaces
(curves) in the heterotic description.  The complex deformations of the
spectral
surface match the
complex deformations of the collection of 7-branes.   Similarly,  
the background
gauge fields inside the 7-branes map to the bundle on the spectral  
surface.
The gauge symmetry enhancement mechanisms are very similar and
the transition of a 5-brane into a finite-size  instanton
is very similar to the instanton --- 3-brane transition.

There is also a similarity in how the matter multiplets appear in both
theories.
In F-theory one expects chiral matter to be produced on the   
intersections
of 7-branes.
On the heterotic side the chiral  matter
multiplets can be expressed as cohomological groups localized to the
intersections of spectral surfaces.
Another  source of (nonchiral) matter in F-theory  is provided by  
the open
strings connecting 7- and 3-branes.
In the heterotic  description, the corresponding multiplets are related
(see section 5) to the  moduli  $H^1 (Hom(\pi^* M,{\cal \V}))$  
responsible for
smoothing out the pointlike instantons (5-branes) into the finite-size
instantons.
It would be very interesting to understand if there is
any  {\it physical} meaning to the analogy between D-branes and spectral
surfaces,
beyond the apparent similarity in the math apparatus.

Another interesting question we only lightly touch upon in section  
3 of this
paper is the appearance of  tensionless strings
\ref\phase{E. Witten, {\it Phase Transitions in M-Theory and  
F-Theory}, Nucl.
Phys.
{\bf B 471} (1996) 195.}\ref\hanahy{A. Hanany and I. R. Klebanov, {\it On
Tensionless Strings in 3+1 Dimensions},
hep-th/9606136.}\mayr\ that should play an
important role in our understanding of various nonperturbative phenomena.
Generically, in four-dimensional compactifications the 7-branes intersect
each other over curves where the singular fiber jumps.
The Calabi-Yau fourfold may require a resolution.
In some cases it is not enough to blow up the
 singular fiber and one also needs to blow up the base.
This leads to a variety of phase transitions.

Clearly, in this paper we just have begun to explore the
four-dimensional F-theory compactifications.
There are plenty of important questions still open,
such as the detailed structure of the map between F-theory --- heterotic
moduli and the  clear understanding of  matter spectrum.
All intricate phenomena known in $N=1$ four-dimensional field theories
should be derivable from F-theory.
One of the real challenges is to  understand the famous Seiberg's duality
\ref\seiberg{N. Seiberg, Nucl. Phys. {\bf B435}  (1995) 129.}.
The first steps in this direction were done in \bjsv\zw.

\newsec{Acknowledgments}

We are grateful to  Ron Donagi, Nikita Nekrasov and Cumrun Vafa
for useful discussions. Our special thanks to Robert Friedman,
John Morgan and Edward Witten for sharing some of  their insights  
with us.
The research of M.~B.~and A.~J.~ was partially supported by the
NSF grant PHY-92-18167, the NSF 1994 NYI award and the  DOE 1994 OJI
award. The research of V.~S.~was supported in part by the NSF grants DMS
93-04580, PHY 9245317 and by Harmon Duncombe Foundation.
The research of T.P. was supported in part by NSF grant
DMS-9500712.

\newsec{Appendix}

Here we list (without proofs) some facts that are used throughout the
paper.

\bigskip

\subsec{Hirzebruch surface $\F_n$ and elliptic Calabi-Yau}

Hirzebruch surface is a $\P^1$ bundle over $\P^1$.
One can think about it as a toric variety. Let $z,w,u$ and $v$ be
coordinates in $C^4$. Define the action of two $U(1)$s, given as follows
$\lambda: (z,w,u,v) \rightarrow (\lambda z,\lambda w, \lambda^n u,v)$
and $\mu: (z,w,u,v) \rightarrow (z, w, \mu u, \mu v)$.  Then the
Hirzebruch surface $\F_n$ is defined as
\eqn\toric{\big(C^4  \setminus  \{{\rm fixed~ set} \}  \big) /
(\lambda, \mu) ~.}
$H^2(\F_n)$ of Hirzerbruch surface is generated by the zero section
$\tilde b$ and a fiber  $\tilde a$.
The intersection pairing of these elements is
${\tilde a}^2=0, {\tilde b}^2=-n$ and $\tilde a \tilde b=1$.

Consider the nonsingular elliptic Calabi-Yau threefold fibered over
$\F_n$.
The fourth cohomology $H^4 (CY_3)$ is three-dimensional
and is generated by $A,B,S$.
We assume that $\pi (A)=\tilde a, \pi (B)=\tilde b$ and $\pi(S)=\F_n$.
The triple intersections are equal to
\eqn\inter{AS^2=-2, ~~BS^2=-2+n, ~~B^2S=-n,~~ ABS=1, ~~S^3=8~,}
all other triple intersections are equal to zero.

\subsec{Generalized Hirzebruch $\F_{nmk}$}

For simplicity we choose the base being the $\P^1$ bundle over the
Hirzebruch surface $\F_n$ (generalized 3-dimensional Hirzebruch ${\bf
F}_{nmk}$).  Let $(z,w,u,v,t,s)$ be the coordinates in $C^6$.
Define three
$U(1)$ actions  as follows
\eqn\uone{\eqalign{
\lambda: (z,w,u,v,s,t) \rightarrow (\lambda z,\lambda w, \lambda^n u,v,
\lambda^m s,t) \cr
\mu : (z,w,u,v,s,t) \rightarrow ( z, w, \mu  u,\mu v, \mu^k s, t) \cr
\nu: (z,w,u,v,s,t) \rightarrow ( z,w,  u,v, \nu s, \nu t) ~.
\cr}}
Then the generalized three-dimensional Hirzebruch is defined as
the quotient
\eqn\toricg{\big(C^6  \setminus  \{{\rm fixed \; set} \}  \big) /
(\lambda,
\mu, \nu)}
For future applications we present here the intersection ring of
$\F_{nmk}$. The ring
is generated by three elements $a,b,c$ satisfying the following relations
\eqn\ring{a^2=0,~~ b^2=-nab, ~~c^2=-kbc-mac~.}
The nonzero intersection pairings are
\eqn\non{abc=1,~~ b^2c=-n, ~~c^3=2km-k^2 n, ~~c^2b=kn-m, ~~c^2a=-k ~.}
In the case of {\it smooth} $CY_4$ one can immediately compute the Euler
character
in terms of some classes of the base \wvs\
\eqn\euil{{1 \over 24} \chi=\int 15c_1^3+ {1 \over 2 } c_1c_2= 732 + 60 k
m - 30 k^2  n ~.}
In the smooth case without any gauge field  inside the 7-branes (trivial
bundle), this number counts the 3-brane (5-brane) anomaly.
In the case when the elliptic fibration has singularities higher
than $I_1$,
the Euler character can be computed using the methods of \klemm.

\bigskip

\subsec{Vector bundles on elliptic fibrations}

\bigskip

\noindent
{\bf Elliptic fibrations.}
An elliptic fibration is a
fibration $\pi : X \rightarrow B$, where $X$ and
$B$ are smooth projective varieties and the $f$ is a flat morphism
whose fibers are connected curves of arithmetic genus one.
Unless stated otherwise we will assume that
the singular fibers of $\pi$ are always reduced and have at most
ordinary double points as singularities. Also we will require that
the fibration $\pi : X \rightarrow B$ possess a section $\sigma :
B \rightarrow X$. In this case $X$ has a natural structure of an
abelian group scheme over $B$ and $\sigma$ is the neutral
element in the Mordell-Weyl group (= the group of global sections).

Denote by $\check{\pi} : \check{X} := {\rm Pic}^{0}(X/B)
\rightarrow B$ the degree zero relative Picard of $\pi$. The general
fibers of $\check{\pi}$ are just the elliptic curves dual to the
corresponding fibers of $\pi$. The existence of $\sigma$
guarantees that as an elliptic fibration
$\check{\pi} : \check{X} \rightarrow B$ is isomorphic to
$\pi : X \rightarrow B$. However, we will keep distinguishing $X$
and $\check{X} $ for the time being  so that we can trace the
sources of the different geometric objects in our construction.

For computational purposes it is convenient to think of the
fibration $\pi : X \rightarrow B$ in terms of its Weierstrass model,
which we proceed to describe.  Put $K_{X/B}$ for the relative
canonical bundle of $\pi$. Its push-forward $\alpha :=
\pi_{*}K_{X/B}$ is a line bundle on $B$ and $X$ sits naturally
as a divisor in ${\bf P}({\cal O}_{B} \oplus \alpha^{\otimes 2}
\oplus \alpha^{\otimes 3})$. Explicitly, there exist sections
$f \in \Gamma(B, \alpha^{\otimes 4})$ and $g \in \Gamma(B,
\alpha^{\otimes 6})$ so that the affine piece of $X$ sitting in
the total space of the vector bundle $\alpha^{\otimes 2}
\oplus \alpha^{\otimes 3}$ is given by the equation
\eqn\weis{
y^{2} = x^{3} + a^{*}f x  + a^{*}g .
}
Here $a : {\bf P}({\cal O}_{B} \oplus \alpha^{\otimes 2}
\oplus \alpha^{\otimes 3}) \rightarrow B$ is the natural projection and
$x$ and $y$ are the tautological sections of the pullbacks of
$\alpha^{\otimes 2}$ and $\alpha^{\otimes 3}$, respectively
\ref\nakayama{N. Nakayama, On Weierstrass models.
Algebraic geometry and
commutative algebra, Vol. II, Kinokuniya, Tokyo, 1988, 405-431. }.
The discriminant locus of $\pi$ is the
divisor of the section $\Delta : = 4f^{3} +27 g^{2} \in \Gamma(B,
\alpha^{\otimes 12})$. For future reference notice that
since $\pi_{*}K_{X} = \alpha\otimes K_{B}$, the variety $X$ will
have a trivial canonical bundle if and only if $\alpha =
K_{B}^{-1}$.

\bigskip
\bigskip

\noindent
{\bf Vector bundles.} We will be interested in instanton bundles
with vanishing first Chern class on elliptic fibrations.  Notice that
if $V \rightarrow X$ is such a bundle, then the restriction of $V$
to any smooth fiber of $\pi$ is a direct sum of indecomposable vector
bundles of degree zero. An indecomposable vector bundle of degree zero
on an elliptic curve is completely determined by its rank $r$ and  
by a line
bundle $\gamma$ of degree zero. More precisely, every such bundle  
is of the
form $E_{r}\otimes \gamma$ where $E_{r}$ is the unique indecomposable
vector bundle of rank $r$ and degree $0$ for which the associated graded
of the Harder-Narasimhan filtration is a direct sum of $r$ copies of the
trivial line bundle ${\cal O}$
\ref\atiyah{M. Atiyah,
Vector bundles over an elliptic curve, Proc. London Math. Soc. vol. 7,
(1957), 414-452.}. It is easy to see that the restriction of a  
general $V$
to the general fiber of $\pi$ will be a direct sum of ${\rm rank}  
(V)$ line
bundles of degree zero. The collection of these line
bundles can be viewed as a collection of points on the dual elliptic
curve and by varying everything over the base we obtain from $V$
a subscheme $\Sigma(V)$ in $\check{X}$ mapping generically finitely  
to $B$
with degree
${\rm rank} (V)$. This subscheme encodes some part of the
geometric information contained in $V$ but is not sufficient for
the reconstruction of $V$. To recover the missing piece of the  
puzzle let us
examine more closely the case when the map $\pi : X \rightarrow B$  
is smooth.
One has the following
\medskip

\noindent
{\bf Proposition 1.} {\it Let $\pi : X \rightarrow B$ be an elliptic
fibration without
singular fibers. The following objects are equivalent
\item{{\rm (i)}} A rank $r$ vector bundle  $V$ on $X$ with
$\det V = {\cal O}_{X}$;
\item{{\rm (ii)}} A pair $(\Sigma, L)$ where $\Sigma
\subset \check{X}$
is a subscheme for which $\check{\pi} : \Sigma \rightarrow B$ is
finite of degree $r$, and $L$ is a rank one sheaf on
$\Sigma$;
}

\medskip

\noindent
Informally we can pass from (i) to (ii) as follows. Take a point $t  
\in B$ and
let $X_{t} = \pi^{-1}(t)$ be the elliptic curve in $X$ sitting over  
$t$ and
$V_{t} = V_{{\mid}X_{t}}$ be the restriction of $V$ to $X_{t}$. As  
we explained
above the bundle $V_{t}$ is a direct sum $V_{t} = E_{k_{i}}\otimes
\alpha_{i}$ with $\alpha_{i} \in  {\rm Pic}^{0}(X_{t})$. The points of
$\Sigma$ sitting over $t$ are the points $\alpha_{i} \in \check{X}_{t}$
($\alpha_{i}$ counts with multiplicity $k_{i}$) and the fiber of $L$ at
the point $\alpha_{i}$ is the vector space $H^{0}(X_{t}, {\rm Hom}(
E_{k_{i}}\otimes \alpha_{i}, V_{t}))$  of all global maps between
$E_{k_{i}}\otimes \alpha_{i}$ and $V_{t}$. Note that in general  
$\alpha_{i}
\neq \alpha_{j}$ and hence  $H^{0}(X_{t}, {\rm Hom}(
E_{k_{i}}\otimes \alpha_{i}, V_{t}))$ is a one-dimensional vector space.

To describe the correspondence of data of type (i) and type (ii) more
rigorously,  consider
the fiber product $X\times_{B} \check{X}$ with the two natural  
projections
$p : X\times_{B} \check{X} \rightarrow X$ and $\check{p} : X\times_{B}
\check{X} \rightarrow \check{X}$. Denote by ${\cal P}$ the relative  
Poincare
bundle on $X\times_{B} \check{X}$ normalized so that the pullback of
${\cal P}$ to $X$ via the zero section $\check{\sigma} : X \rightarrow
X\times_{B} \check{X}$ is ${\cal O}_{X}$ and  the pullback of
${\cal P}$ to $\check{X}$ via the zero section $\sigma : \check{X}  
\rightarrow
X\times_{B} \check{X}$ is ${\cal O}_{\check{X}}$. Recall that  
${\cal P}$ is
uniquely characterized by the normalization condition and by the  
property:
for every $t \in B$ and every $\alpha \in \check{X}_{t} =  
\check{p}^{-1}(t)
= {\rm Pic}^{0}(X_{t})$ there is an isomorphism ${\cal  
P}_{{\mid}X_{t}\times
\{ \alpha \}} \cong \alpha$. Now we are ready to formalize the passage
between the data (i) and (ii).

Starting with a vector bundle $V$ on $X$ with trivial determinant,  
we can form
the sheaf ${\cal F}(V) := \check{p}_{*}(p^{*}V\otimes {\cal P}^{-1})$ on
$\check{X}$. By construction ${\cal F}(V)$ is a
torsion sheaf on $\check{X}$ supported at the set of points $\alpha \in
\check{X}$ that have the property $\dim H^{0}(X_{\check{\pi}(\alpha)},
\alpha^{-1}\otimes V_{\check{\pi}(\alpha)}) \neq 0$. In
particular the support $\Sigma(V)$ of
${\cal F}(V)$ is a divisor in $\check{X}$ that maps $r:1$ to the  
base $B$.
Alternatively ${\cal F}(V)$ can be thought of as the extension by  
zero of a
sheaf $L(V)$ on $\Sigma(V)$ and it is not hard to check that $L(V)$  
must have
rank one.
Conversely, if we start with a pair $(\Sigma, L)$ we can construct
a vector bundle $V(\Sigma,L)$ of rank $r$ on $X$ in the following way:
The
fiber product $Y := X\times_{B} \Sigma$ is a smooth elliptic  
fibration over
$\Sigma$ via the natural projection $p_{\Sigma} : Y
\rightarrow \Sigma$. Put $p_{X} : Y\rightarrow X$ for the
projection on $X$ and define $V(\Sigma, L) =  
p_{X*}(p_{\Sigma}^{*}L\otimes
{\cal P} \otimes \omega_{Y/X}^{-1})$ where  as before ${\cal P}$ is the
(restriction of) the Poincare bundle from $X\times_{B} \check{X}$ and
$\omega_{Y/X}$ is the relative dualizing sheaf of the map $p_{X} : Y
\rightarrow X$. It is not hard to convince oneself that the two  
assignments
$V \mapsto (\Sigma(V),L(V))$ and $(\Sigma,L) \mapsto V(\Sigma,L)$  are
inverse to each other. Also note, that in general position $\Sigma$  
will be a
smooth cover and $L$ will be a line bundle on $\Sigma$.
In that case $Y$
is also smooth and by the Hurwitz formula $\omega_{Y/X}^{-1} = {\cal
O}_{Y}(-R)$, with $R \subset Y$ being the ramification divisor of the
projection
$p_{X} : Y \rightarrow X$.

\medskip

When we allow singular fibers in the elliptic fibration $\pi : X  
\rightarrow
B$ the above simple correspondence between the data (i) and (ii) does not
hold literally even when we are in general position. It turns out that
the smoothness of $\Sigma$ does not in general imply the smoothness of
$Y$ and that it is necessary to modify the assignment $(\Sigma,L) \mapsto
V(\Sigma,L)$ along the singular locus of $Y$.
Instead of discussing the
necessary modifications in full generality, we will briefly explain below
what needs to be done in the specific situations when $X$ is $K3$ surface
or a Calabi-Yau 3-fold.

\medskip

\noindent
Among other things this description of vector bundles on $X$ leads to a
peculiar compactification of the moduli space which is obtained as
follows:
Fix a set of cohomology classes  $\unc \in \oplus_{i \geq 2}
(H^{i,i}(X)\cap H^{2i}(X,{\bf Z}))$ on $X$. Fix an ample line
bundle $H$ on
$X$ and denote by $M_{X}(r,\unc)$ the moduli space of rank $r$ bundles on
$X$ with Chern classes $\unc$ that are Gieseker semistable with
respect to
$H$. The polarization $H$ induces a canonical polarization $\tilde{H}$ on
the
fiber product $X\times_{B} \check{X}$ and the $H$-stability condition on
a bundle $V$ is equivalent to the $\tilde{H}$-stability of $L$ considered
as
a torsion sheaf on $X\times_{B} \check{X}$ supported on the divisor
$X\times_{B} \Sigma$. The Hilbert polynomial $p$ of this torsion
sheaf can
be
calculated entirely in terms of $\unc$ and so we can identify
$M_{X}(r,\unc)$
with a Zariski open subset in the moduli of sheaves
$M_{X\times_{B} \check{X}}(p)$.
The structure of the latter is rather
simple.
The morphism assigning to a sheaf its support realizes $M_{X\times_{B}
\check{X}}(p)$ as a fibration over the set of all $\Sigma$'s whose fibers
are compactified Picard varieties.
It is easy to see that for a fixed $\unc$
the various divisors $\Sigma$ are all linearly equivalent and so the
base of this fibration is a projective space.

\subsec{The  $K3$ case}

Let us examine the case when $X$  is a $K3$ surface in more
details. In this case the base $B$ is the projective line. We will
assume that $X$ is generic in the sense that $\pi$ has exactly
24 singular fibers. This situation has the advantage that at least
for a general $V$ the branch points of $\Sigma(V)$ will be different from
the discriminant of $\pi : X \rightarrow B$ and so the fiber product
$Y = X\times_{B}\Sigma(V)$ will be smooth. This allows us to use the
above procedure for passing between $V$ and $(\Sigma,L)$ without any
further modifications. Let $S_{X}, F_{X}$ be the classes of the zero
section and the fiber, respectively.  It is known that
$F_{X}^{2} = 0$,  $F_{X}S_{X} = 1$ and $S_{X}^{2} = -2$ and
that for a general  such $X$  we have ${\rm Pic}(X) = {\bf
Z}S_{X}\oplus {\bf Z}F_{X}$. Similarly we have
${\rm Pic}(\check{X}) = {\bf Z}S_{\check{X}}\oplus {\bf Z}
F_{\check{X}}$.

Using these bases it is not hard to express the numerical invariants of
$(\Sigma,L)$ in terms of the numerical invariants of $V$. If $V$ is  
a rank
$r$ vector bundle on $X$ with trivial first Chern class,
then by construction $\Sigma(V) = rS_{\check{X}} + kF_{\check{V}}$.  
To find
the coefficient $k$ consider the intersection $S_{\check{X}}\cdot  
\Sigma(V)$.
For $V$ in general position, $S_{\check{X}}\cdot \Sigma(V)$ consists of
$k - 2r$ distinct points\foot{Notice that since $S_{\check{X}}$ and
$\Sigma(V)$ are both effective, $k > 2r$ is a necessary condition for the
existence of the bundle $V$.} on $S_{\check{X}} \subset \check{X}$.
A point
$\alpha \in S_{\check{X}}\cap \Sigma(V)$ corresponds to a copy of the
trivial line bundle appearing as a direct summand in
$V_{\check{\pi}(\alpha)}$.
Therefore $k - 2r = \dim H^{0}(B, \pi_{*}V) =
\dim H^{0}(B, R^{1}\pi_{*}V)$.
Furthermore, the fact that $\pi_{*}V$ is a
torsion sheaf on the curve $B$ and the Lerray-Serre spectral  
sequence imply
that $\dim H^{1}(X, V) = \dim H^{0}(B, R^{1}\pi_{*}V) = k -2r$. If $V$ is
stable on $X$, then $H^{0}(X, V) = 0$ and by duality $H^{2}(X,V) = 0$.
Thus $k - 2r = \dim H^{1}(X, V) = - \chi(X, V)$ which can be calculated
by the Hirzebruch-Riemann-Roch formula. We have
$$
2r - k = \chi(X,V) = [ch(V)td(T_{X})]_{2} = [(r - c_{2}(V)t^{2})(1 +
2t^{2})]_{2} = 2r - c_{2}(V),
$$
and so $k = c_{2}(V)$. To calculate the degree of the line bundle $L(V)$
in terms of $V$ we just have to notice that $\sigma^{*}(V)$ is  
going to be
the pushforward of the line bundle on $\Sigma(V)$ which is the  
restriction
of $p_{\Sigma}^{*}L(V)\otimes {\cal P} \otimes \omega_{Y/X}^{-1}$ to
the zero section of the elliptic fibration $p_{\Sigma} : Y \rightarrow
\Sigma$.
By the normalization condition on the Poincare bundle we know that
${\cal P}$ restricts to the trivial line bundle on this section.
Also, since
$Y$ is the fiber product of $\Sigma$ and $X$, it follows that the  
ramification
divisor of the covering $p_{X} : Y \rightarrow X$ is the pull-back of the
ramification divisor of $\check{\pi} : \Sigma(V) \rightarrow B$, which,
combined with the condition $\deg \sigma^{*}(V) = 0$, yields $\deg L(V) =
-1/2 \deg K_{\Sigma(V)/B}$. To summarize:

\medskip

\noindent
{\bf Claim 1.} {\it Let $V \rightarrow X$ be a rank $r$ vector
bundle
with $\det (V) = {\cal O}_{X}$. Let $(\Sigma, L)$ be the pair  
corresponding
to $V$.  Then $\Sigma = rS_{\check{X}} + c_{2}(V)F_{\check{X}}$ and
$\deg L = -(r + g -1)$ where $g = g(\Sigma) = rc_{2}(V) - r^{2} + 1$.
}

\bigskip

In these terms it is easy to describe the moduli space
$M_{X}(r,k)$ of rank $r$ vector bundles on $X$ with trivial first Chern
class and second Chern number $k$.
Fix a smooth $\Sigma \subset
\check{X}$ which is linearly equivalent to $rS_{\check{X}} +
kF_{\check{X}}$.
Let $M_{\Sigma} \subset M_{X}(r,k)$ be the
subvariety, parameterizing vector bundles giving rise to $\Sigma$.
The fibration $p_{\Sigma} : Y \rightarrow
\Sigma$ has two natural sections $S_{Y}$ and $T_{Y}$.
The
section $S_{Y}$ is the pull-back of $S_{X}$ via $p_{X}$ and
$T_{Y}$ is the graph of the embedding $\Sigma \subset X$
obtained from the identification $X \cong  \check{X}$.
It is straightforward
to calculate the intersections of $S_{Y}$ and $T_{Y}$ on $Y$.
We have
$S_{Y}^{2} = -2r, T_{Y}^{2} = -2r, S_{Y}T_{Y} = k - 2r$.
Also for a general
$Y$ one has ${\rm Pic}(Y) = {\rm Pic}(\Sigma)\oplus {\bf Z}S_{Y} \oplus
{\bf Z}T_{Y}$.
For the Poincare bundle ${\cal P}$ one calculates
${\cal P} = {\cal O}_{Y}(S_{Y} - T_{Y})\otimes p_{\Sigma}^{*}  
\rho$, where
$\rho$ is line bundle on $\Sigma$ of degree $-c_{2}(V)$. This  
combined with
Proposition~1 gives

\medskip

\noindent
{\bf Claim 2.} {\it Let $d = c_{2}(V) + r + g -1$. The natural map
$\varphi_{\Sigma}: {\rm Pic}^{d}(\Sigma) \rightarrow M_{\Sigma}$ given by
$\xi \mapsto p_{X*}(p_{\Sigma}^{*}\xi\otimes {\cal O}_{Y}(S_{Y} -  
T_{Y}))$
is an isomorphism.}

\bigskip

As a corollary we immediately obtain

\medskip

\noindent
{\bf Corollary 1.} {\it $M_{X}(r,k)$ is birationally isomorphic
to the total space
of the family of Jacobians of degree $k + r  + g - 1$ (equivalently
$-(r + g - 1)$) of the curves in the linear system $\mid  
rS_{\check{X}} + k
F_{\check{X}} \mid $. In particular the smooth part of $M_{X}(r,k)$ is a
hyperk\"ahler manifold which is also a completely integrable
Hamiltonian system.}

\subsec{The Calabi-Yau case}

Suppose now that $B\cong {\bf F}_{n}$ is a Hirzebruch surface and
that $X$
is a three-dimensional Calabi-Yau manifold.
We know that ${\rm Pic}(B) =
{\bf Z}\tilde{a}\oplus {\bf Z}\tilde{b}$,  where $\tilde{a}$ is the fiber
of the Hirzebruch surface $B$
and $\tilde{b}$ is the infinity section. We have $\tilde{a}^{2} = 0$,
$\tilde{b}^{2} = -n$ and $\tilde{a}\tilde{b} = 1$.
The Chow ring of a
generic $X$ of this type is generated by the three divisor classes
$A_{X} :=
\pi^{*}\tilde{a}$, $B_{X} := \pi^{*}\tilde{b}$ and
$S_{X} = \sigma(B)$, with relations $A_{X}^{2} = B_{X}^{2}
= 0$ and
the ones given by the formulas \inter. In particular the Picard
group of $X$
is freely generated by $A_{X}, B_{X}, S_{X}$ as
an abelian group
and the integral part of $H^{2,2}(X)$
is freely generated by the curves $A_{X}B_{X}$, $B_{X}S_{X}$
and $A_{X}S_{X}$.
Similarly we have classes $A_{\check{X}}$,
$B_{\check{X}}$ and $S_{\check{X}}$ for $\check{X}$.
Thus we can
write $c_{2}(V) = c_{2}(V)_{A B} A_{X}B_{X} +
c_{2}(V)_{A S} A_{X} S_{X} +
c_{2}(V)_{B S} B_{X} S_{X}$
for any vector bundle $V$ on $X$. As in the construction
discussed in Proposition~1 we can form the torsion sheaf
$\check{p}_{*}(p^{*}V\otimes {\cal P}^{-1})$. Its support  
$\Sigma(V)$ will be
a surface in $\check{X}$ for which the map $\check{\pi} : \Sigma(V)
\rightarrow B$ is generically finite of degree $r$.
To recover the numerical
invariants of $(\Sigma(V),L(V))$ in terms of those of $V$, we just  
have to
notice that the general members of the linear
systems ${\mid}A_{X}{\mid}$ and ${\mid}B_{X}
+ n/2 A_{X}{\mid}$ are smooth elliptic $K3$ surfaces sitting in $X$.
After
restricting to those and applying what we already know about the $K3$
case we obtain $\Sigma(V) = r S_{\check{X}} + c_{2}(V)_{A S}
A_{\check{X}} + c_{2}(V)_{B S} B_{\check{X}}$.
The information about the
component $c_{2}(V)_{AB}$ of $c_{2}(V)$ can also be read off from  
the pair
$(\Sigma, L)$.
Indeed, since $c_{2}(V)_{AS}$ and $c_{2}(V)_{BS}$ are
determined by $\Sigma$, it suffices to compute the intersection $c_{2}(V)
\cdot S_{X} = c_{2}(V_{\mid S_{X}})$ in terms of $L$ and $\Sigma$. On the
other hand by construction we have $V_{\mid S_{X}} = \pi_{*}(L\otimes
\omega_{\Sigma/B}^{-1})$.
Put $M := L\otimes\omega_{\Sigma/B}^{-1}$.
The Grothendieck-Riemann-Roch formula for the finite map
$\pi : \Sigma \rightarrow B$ reads
$$
ch(\pi_{*}M)td(T_{B}) = \pi_{*}(ch(M)td(T_{\Sigma}))
$$
and in combination with the condition $c_{1}(V) = 0$ and the Riemann-Roch
theorem for the line bundle $M$ on the surface $\Sigma$ this yields
$$
c_{2}(V)\cdot S_{X} = r\cdot td_{2}(T_{B}) + (K_{\Sigma}\cdot M -  
M^{2})/2 -
td_{2}(T_{\Sigma}) = r\cdot td_{2}(T_{B}) - \chi(M) + \chi({\cal
O}_{\Sigma}) - td_{2}(T_{\Sigma}).
$$
It is also straightforward to check that for the Hirzebruch surface  
$B$ one
has $td_{2}(T_{B}) = 1$. Since the Hirzebruch-Riemann-Roch formula on
$\Sigma$ gives  $\chi(\Sigma, {\cal O}_{\Sigma}) =  
td_{2}(T_{\Sigma})$, we get
summarily the following

\bigskip

\noindent
{\bf Claim 3.} {\it Let $V \rightarrow X$ be a rank $r$ vector bundle
with $\det (V) = {\cal O}_{X}$. Let $(\Sigma(V), L(V))$ be the pair
corresponding to $V$.  Then
$\Sigma(V) = r S_{\check{X}} + c_{2}(V)_{A S}
A_{\check{X}} + c_{2}(V)_{B S} B_{\check{X}}$ in ${\rm
Pic}(\check{X})$
and $c_{2}(V)\cdot S_{X} = r - \chi(\Sigma(V), L(V)\otimes
\omega_{\Sigma(V)/B}^{-1})$.
}

\bigskip

\noindent
In order to recover the bundle $V$ from the pair $(\Sigma(V),L(V))$,
we have to modify slightly the construction from Proposition~1.
The
modification is forced by the fact that even when $\Sigma(V)$ is smooth
the fibered product $X\times_{B} \Sigma(V)$ will be singular since
the branch divisor of $\Sigma(V) \rightarrow B$ will always intersect the
discriminant of $\pi : X \rightarrow B$, which is ample.
The singularities
of $X\times_{B} \Sigma(V)$ occur at the intersection points of the
branch and the discriminant divisors and are therefore isolated. If
$\nu : Y \rightarrow X\times_{B} \Sigma(V)$ denotes a resolution of these
singularities and $p_{\Sigma} : Y \rightarrow \Sigma(V)$ and $p_{X} :
Y \rightarrow X$ are the natural projections, one can check that
$V$ is isomorphic to the push-forward $p_{X*}{\cal L}(V)$ of a suitable
rank one sheaf ${\cal L}(V) \to Y$.
The sheaf ${\cal L}(V)$ can be
reconstructed from $L(V)$ as ${\cal L} = p_{\Sigma}^{*}L(V)\otimes
\nu^{*}{\cal P}
\otimes \omega^{-1}_{Y/X} \otimes {\cal O}_{Y}(\ell E)$, where $E \subset
Y$ is the exceptional divisor of $\nu$.
The integer $\ell$ is completely
determined by (and determines) the third Chern class of $V$.

\medskip

\noindent
It can be checked that the condition that the linear system
${\mid}r S_{\check{X}} + c_{2}(V)_{A S}
A_{\check{X}} + c_{2}(V)_{B S} B_{\check{X}}{\mid}$ contains an
effective
divisor, implies that the line bundle
${\cal O}_{\check{X}}
(r S_{\check{X}} + c_{2}(V)_{A S} A_{\check{X}} +
c_{2}(V)_{B S} B_{\check{X}})$ is actually ample on
$\check{X}$. By Bertini's theorem the general spectral surface
$\Sigma$ will
be smooth and connected. Moreover for such a surface the Lefschetz
hyperplane
section theorem gives $H^{1}(\Sigma, {\cal O}_{\Sigma}) = 0$ and
$H^{1,1}(\check{X}) \subset H^{1,1}(\Sigma)$. Let $\unc$ be the
Chern classes
of $V$. For a fixed $(\Sigma, {\cal L})$ the K\"{u}nneth formula
applied to
$X\times_{B} \Sigma$ shows that ${\cal L}$ and $L$ have the same
number of
moduli. Therefore the support map $M_{X}(r,\unc) \rightarrow {\mid}
 r S_{\check{X}} + c_{2}(V)_{A S}
A_{\check{X}} + c_{2}(V)_{B S} B_{\check{X}} {\mid}$ is surjective and
generically finite.

\bigskip

\noindent
{\bf Remarks.}

\smallskip

1. In contrast with the $K3$ case, the moduli space $M_{X}(r,\unc)$ may
be reducible and may have components of different dimension. It can also
happen that the support map contracts whole components of
$M_{X}(r,\unc)$.
Examples like that can be easily constructed by taking direct sums of
bundles on $X$ with pull-backs of bundles on $B$.

\smallskip

2. By degenerating $X$ to a double generalized Hirzebruch,
it can be shown that for the general pair $(X,\Sigma)$ the only divisor
classes on $\Sigma$ are the restrictions of $A_{\check{X}}$,
$B_{\check{X}}$ and $S_{\check{X}}$. In particular ${\cal L}$ is a
linear combination of the exceptional divisor $E$ and of the
strict transforms of the zero section of
$X\times_{B} \Sigma \rightarrow \Sigma$, the divisor ${\cal T} \subset
X\times_{B}\Sigma$
corresponding to the graph of the embedding $\Sigma \subset X$, and
the pull-backs of $A_{\check{X}\mid \Sigma}$, $B_{\check{X}\mid \Sigma}$
and
$S_{\check{X}\mid \Sigma}$ to $X\times_{B} \Sigma$. The six
coefficients of ${\cal L}$ in this basis are not independent. There are
relations between them coming from the identification
$p_{X*}{\cal L} = V$ and the condition $c_{1}(V) = 0$ and from fixing
$c_{3}(V)$ and $c_{2}(V)\cdot S_{X}$.

\listrefs
\end

This
ambiguity
is characterized by two integer parameters. The Picard number for
the spectral
surface $\S$ is equal to $3$ and therefore all line bundles are
classified by
three numbers
-- the components of $c_1(L)$. The line bundle $L$ does not
have enough
integer valued parameters to describe all vector bundles. The
Picard number for
the fibered product is equal to $5$
and therefore all lines bundles are classified by five numbers
-- the components of $c_1({\cal L})$.  Now we have enough
integer valued parameters to classify the vector bundles, namely  
the three
components of $\Sigma$ and the components of $c_{1}({\cal L})$. The
condition
that $c_{1}(V) = 0 $ imposes three restrictions on $\Sigma$ and  
$c_1({\cal
L})$ leading to five independent parameters. One can find a mathematical
discussion in Appendix.

be able
is worth